\documentclass[journal,10pt, final]{IEEEtran}  
\usepackage{amsmath}
\usepackage{bm}
\usepackage{amsfonts}
\usepackage{amssymb}
\usepackage{lineno}
\usepackage{graphics} 
\usepackage{mathtools}
\usepackage{cite}
\usepackage{multirow}
\usepackage{threeparttable}
\usepackage{booktabs}
\usepackage{xr}

\DeclareMathOperator*{\argmin}{arg\,min}

\usepackage{algorithmicx}
\usepackage{algorithm}
\usepackage[noend]{algpseudocode}   
\algnewcommand\algorithmicinput{\textbf{Input:}}
\algnewcommand\Input{\item[\algorithmicinput]}

\algnewcommand\algorithmicoutput{\textbf{Output:}}
\algnewcommand\Output{\item[\algorithmicoutput]}

\algnewcommand{\LineComment}[1]{\Statex \(\triangledown \) #1}  

\algnewcommand{\IfThenElse}[3]{
	\State \algorithmicif\ #1\ \algorithmicthen\ #2\ \algorithmicelse\ #3 \algorithmicend}
\algnewcommand{\IfThen}[2]{\State \algorithmicif\ #1\ \algorithmicthen\ #2 \algorithmicend}

\DeclareGraphicsExtensions{.pdf,.jpeg,.png,.tiff}

\newtheorem{definition}{Definition}

\newtheorem{remark}{Remark}
\newtheorem{lemma}{Lemma}
\newtheorem{proposition}{Proposition}
\newtheorem{theorem}{Theorem}
\newtheorem{corollary}{Corollary}
\newtheorem{problem}{Problem}
\newtheorem{example}{Example}

\usepackage{mathtools}
\newcommand{\deq}{\vcentcolon=}

\usepackage{hyperref}
\usepackage{xcolor}
\hypersetup{
	colorlinks,
	linkcolor={red!50!black},
	citecolor={blue!50!black},
	urlcolor={blue!80!black}
}
\newcommand{\cL}{\mathcal L}
\newcommand{\cR}{\mathcal R}
\newcommand{\cC}{\mathcal C}
\newcommand{\cE}{\mathcal E}
\newcommand{\bbN}{\mathbb N}
\newcommand{\bbR}{\mathbb R}
\newcommand{\bu}{\bm u}
\newcommand{\ba}{\bm{a}}
\newcommand{\bsg}{\bm \sigma}
\newcommand{\DN}{\Delta_N}
\newcommand{\Dz}{\Delta_z}
\newcommand{\dN}{\delta_N}
\newcommand{\DM}{\Delta_M}
\newcommand{\dM}{\delta_M}
\newcommand{\dz}{\delta_z}
\newcommand{\bw}{\bar{w}}

\newcommand{\pt}{p_{\textrm{t}}}
\newcommand{\Go}{G_{\textrm{o}}}
\newcommand{\ps}{p_{\textrm{s}}}

\newcommand{\rv}[1]{\textcolor{black}{#1}}

\begin{document}
%
\title{Infinite-Horizon Optimal Control of Switched Boolean Control Networks with Average Cost: An Efficient Graph-Theoretical Approach}
%
%
%
\author{Shuhua~Gao,
	Changkai~Sun,
	Cheng~Xiang,~\IEEEmembership{Member,~IEEE,}
	Kairong~Qin, ~\IEEEmembership{Member,~IEEE,}\\
	and Tong Heng~Lee,~\IEEEmembership{Member,~IEEE}
	\thanks{Shuhua Gao, Cheng Xiang, and Tong Heng Lee are with the Department
		of Electrical \& Computer Engineering,  National University of Singapore, 119077 Singapore, (e-mail: elegaos@nus.edu.sg; elexc@nus.edu.sg; eleleeth@nus.edu.sg).}
	\thanks{Changkai Sun is with the Research \& Educational Center for the Control Engineering of Translational Precision Medicine (RECCE-TPM), School of Biomedical Engineering, and Kairong Qin is with the School of Optoelectronic Engineering and Instrumentation Science, Dalian University of Technology, Dalian 116024, China, (e-mail: sunck2@dlut.edu.cn, krqin@dlut.edu.cn).}
}
%
%

\markboth{To appear in IEEE Transactions on Cybernetics (DOI: 10.1109/TCYB.2020.3003552)}
{Shell \MakeLowercase{\textit{et al.}}: Bare Demo of IEEEtran.cls for IEEE Journals}
%



\maketitle

\begin{abstract}	
This study investigates the infinite-horizon optimal control problem for switched Boolean control networks with an average-cost criterion. A primary challenge of this problem is the prohibitively high computational cost when dealing with large-scale networks. We attempt to develop a more efficient approach from a novel graph-theoretical perspective. First, a weighted directed graph structure called the \textit{optimal state transition graph} (OSTG) is established, whose edges encode the optimal action for each admissible state transition between states reachable from a given initial state subject to various constraints.  Then, we reduce the infinite-horizon optimal control problem into a minimum mean cycle (MMC) problem in the OSTG. Finally, we develop an algorithm that can quickly find a particular MMC by resorting to Karp's algorithm in graph theory and construct an optimal switching-control law based on state feedback. Time complexity analysis shows that our algorithm, \rv{albeit still running in exponential time}, can outperform all existing methods in terms of time efficiency. A 16-state-3-input signaling network in leukemia is used as a benchmark to test its effectiveness. Results show that the proposed graph-theoretical approach is much more computationally efficient and can reduce the running time dramatically: it runs hundreds or even thousands of times faster than existing methods. The Python implementation of the algorithm is available at \url{https://github.com/ShuhuaGao/sbcn_mmc}.
\end{abstract}


\IEEEpeerreviewmaketitle

\section{Introduction} \label{sec: intro}
%
%
%
%
\IEEEPARstart{B}{oolean} networks (BNs), proposed by Kauffman \cite{kauffman1969metabolic}, represent a special class of discrete-time logical systems with binary state variables. The most important application of BNs is to model complex biomolecular networks, especially the gene regulatory networks (GRNs) \cite{wang2012boolean}. The binary state of a gene in a GRN indicates whether this gene is expressed, and each gene's state is updated by a specific Boolean function characterizing the regulatory interaction among genes. A Boolean network involving exogenous inputs, which may indicate external interventions in a therapeutic context \cite{pal2006optimal,li2015controllability}, is commonly termed a Boolean control network (BCN). 

In recent years, the semi-tensor product (STP)  of matrices, developed by Cheng et al. \cite{cheng2010linear,zhao2010input},  has revived the studies on BCNs by formalizing an algebraic state-space representation (ASSR). Under this framework, a variety of well-established techniques in conventional control theory can be adapted to handle similar problems for BCNs. To date, many control-theoretical problems related with BCNs have been investigated using this new toolset, including controllability and observability \cite{zhao2010input,laschov2013observability}, stabilization \cite{cheng2011stability}, pinning control \cite{zhong2018pinning}, and optimal control \cite{zhao2010optimal,zhao2011floyd,fornasini2013optimal,cheng2014optimal}, just to name a few. 

A variant of BCNs drawing much attention is the switched Boolean control network (SBCN), an analogy to the traditional switched system \cite{sun2005analysis}, whose dynamics is governed by multiple network models of different structures and (or) logical rules. SBCNs have a solid biological foundation such as the four-stage growth and division of eukaryotic cells that exhibit different dynamics \cite{hatzimanikatis1999mathematical}. To describe these possibly time-varying BNs,  refinements such as the probabilistic Boolean networks (PBNs) \cite{pal2006optimal}, and the asynchronous updating scheme \cite{wang2012boolean} have been proposed, all of which can be viewed as an SBCN with a particular (possibly nondeterministic) switching law. Based on the ASSR of SBCNs, some interesting control-theoretical problems have been recently addressed, e.g., stability analysis \cite{li2012reachability}, stabilizable controller synthesis \cite{yu2018stabilizability}, and its application to stability analysis of Boolean control networks under aperiodic sampled-data control (ASDC) \cite{lu2018stabilization, sun2019switching}.

Optimal control aims to find a control law to optimize a given performance criterion. One medical application of optimal control of BCNs is to design the best therapeutic intervention strategy \cite{pal2006optimal}. Finite-horizon optimal control of BCNs has been widely studied, e.g., \cite{fornasini2013optimal} and \cite{laschov2010maximum, li2013minimum, laschov2013minimum,zhu2018optimal}. This study focuses on infinite-horizon optimal control (IHOC) of SBCNs with average cost, which has been previously attempted in \cite{zhao2010optimal,zhao2011floyd,fornasini2013optimal,wu2019optimal} towards BCNs and \cite{li2014optimal} for SBCNs. Specifically, the IHOC problem with average-cost criteria was first addressed in \cite{zhao2010optimal} by enumerating all cycles in the input-state space of a BCN, and its efficiency was enhanced afterward in \cite{zhao2011floyd} using a Floyd-like algorithm. After that, the same problem for BCNs was studied in \cite{fornasini2013optimal} and \cite{wu2019optimal} with value iteration and policy iteration based approaches, respectively. By contrast, the IHOC of SBCNs with average cost was only investigated in \cite{li2014optimal} using a simple variant of the Floyd-like algorithm \cite{zhao2011floyd}. Besides, another common class of IHOC problems for BCNs with discounted cost has been considered in \cite{cheng2014optimal,cheng2015receding,zhu2018optimal}. 

\rv{As emphasized in the most recent work \cite{wu2019optimal}, the primary challenge in IHOC of BCNs is its high computational cost, which can result in computational intractability in case of large networks. More generally, 
the intensive computational burden of most BCN-related tasks, not limited to optimal control, is mainly caused by the so-called \textit{state space explosion}: a BCN with $ n $ variables has a total of $N \deq 2^n $ states. This issue has been emphasized in various studies like \cite{ zhao2010input,zhao2010optimal,laschov2013observability,liang2017improved,wu2019optimal}.  For example, the NP-hardness to examine the controllability and observability of BCNs has been proved in \cite{akutsu2007control} and \cite{laschov2013observability} respectively. Consequently, most algorithms reviewed above run in  polynomial time of $ N $ of high degree and can only handle tiny networks. Nonetheless, there is still room for computational efficiency improvement by reducing the degree of this polynomial in $ N $. Thus, the primary goal of this study is to design more efficient algorithms with reduced time complexity for SBCNs by constructing both control inputs and switching signals based on state feedback.} 

\rv{To this end, we note a distinctive property of a BCN (SBCN): its state space and control space are both finite, and its state transitions are deterministic. This property allows us to encode the full dynamics of a BCN into a graph, known as the \textit{state transition graph} (STG). We thus may improve computational efficiency for control-theoretical problems of BCNs by utilizing algorithms in graph theory. One example is the graphical description of BCN stabilization acquired by two in-tree search algorithms \cite{liang2017algorithms}. Another example is the modified controllability criterion of BCNs using the Warshall algorithm \cite{liang2017improved} and Tarjan's algorithm \cite{zhu2018further}. These pioneering research reveals the potential to accelerate BCN-related algorithms by combing the ASSR and graph theory, which motivates our study here. As far as we know, this paper presents the first attempt to solve the average-cost IHOC problem in a graph-theoretical framework with high efficiency.}

The main contributions are four folds. (i) We establish a graph structure called the \textit{optimal state transition graph} (OSTG), which depicts the optimal action for the transition between each pair of connected states and can handle state constraints as well as state-dependent control and switching constraints elegantly. (ii) The IHOC problem is reduced to a minimum mean cycle (MMC) problem in the OSTG. We resort to Karp's method in graph theory for fast MMC search and develop a novel algorithm with supreme time efficiency. A state-feedback control and switching law are constructed by our algorithm to achieve optimal control. (iii) Our graph-theoretical approach reduces the time complexity from the state-of-the-art $ O(MN + N^4) $ to $ O(MN^2) $, where $ M \deq 2^m $ and $ m $ is the number of control inputs. The effectiveness and efficiency of the proposed approach have been verified with a 16-node network involved in leukemia. Results show that our approach can outperform all existing methods in terms of computational efficiency with a significant advantage: it runs hundreds or even thousands of times faster.  \rv{(iv) Additionally, the proposed approach can be easily extended to solve the optimal control problem for all initial states instead of a particular one without losing its efficiency advantage.}

The remainder of this paper is organized as follows. In Section \ref{sec: pre}, we introduce the ASSR of BCNs and present some fundamental concepts in graph theory. The IHOC problem of SBCNs with average cost is formulated in Section \ref{sec: problem}. The key data structure of our algorithms, i.e., the optimal state transition graph, is presented in Section \ref{sec: ostg}. We design the algorithms for IHOC of SBCNs with average cost in Section \ref{sec: algrithms} and compare the time complexity of our approach with that of existing methods in Section \ref{sec: time complexity}. The performance of these methods is benchmarked in Section \ref{sec: benchmark} using a 16-state and 3-input network. Finally, Section \ref{sec: conclusion} concludes this study.

\section{Preliminaries} \label{sec: pre}
\subsection{Notations}
The following notations, mainly adopted from \cite{cheng2010linear} and \cite{zhao2010input} for the STP and the ASSR of BCNs, are used.$ |S| $ denotes the size (i.e., cardinality) of a set $ S $. $ \bbR $, $ \bbN $, and $ \bbN^+ $ denote the sets of real numbers, nonnegative integers, and positive integers respectively. $\mathcal{M}_{p\times q}$ denotes the set of all $p\times q$ matrices. 
	$ \textrm{Col}_i(A) $ denotes $ i $-th column of a matrix $ A $, and $ A_{ij} $ denotes the $ (i, j) $-th entry of the matrix $ A $.
 Set $ \delta_n^i \deq \textrm{Col}_i(I_n) $, where $ I_n $ is the $ n $-dimensional identity matrix. Let  $\Delta_n = \{ \delta_n^i | i = 1, 2, \cdots, n \}$, and set $ \Delta \deq \Delta_2 $.  
 A $ n\times q $ matrix  $A =[ \delta_{n}^{i_1}\; \delta_{n}^{i_2}\; \cdots\; \delta_{n}^{i_q} ] $ with $\delta_n^{i_k}  \in \Delta_n, 1 \le k \le q$, is called a \textit{logical matrix}. Let $\mathcal{L}_{n\times q} $ denote the set of all $ n\times q $ logical matrices.
 A shorthand notation for a matrix $ A =[ \delta_{n}^{i_1}\; \delta_{n}^{i_2}\; \cdots\; \delta_{n}^{i_q} ]$ is $ A = \delta_{n}[i_1, i_2, \cdots, i_q] $. This condensed form applies to a set, a sequence, and a path as well.
 A matrix $ A \in  \mathcal{M}_{n\times mn}$ can be rewritten into a block form $ A = [\text{Blk}_1(A)\; \text{Blk}_2(A)\; \cdots\; \text{Blk}_m(A)] $, where $ \text{Blk}_i(A) \in \mathcal{M}_{n \times n}$ is the $ i $-th square block of $ A$.
 Common logical operators \cite{cheng2010linear} are listed as follows. $\land$: conjunction; $\lor$: disjunction; $ \lnot $: negation; $ \leftrightarrow $: equivalence;  $ \oplus $: exclusive or; and $ \to $: implication.

\subsection{Algebraic Form of SBCNs} \label{sec: ASSR}
\begin{definition}\cite{zhao2010optimal}
	The semi-tensor product (STP) of two matrices $ A \in   \mathcal{M}_{m\times n}$ and $ B \in   \mathcal{M}_{p\times q}$ is defined by
	\begin{equation*}
	A \ltimes B = (A \otimes I_{\frac{s}{n}})(B \otimes I_{\frac{s}{p}}),
	\end{equation*}
	where $\otimes$ denotes the Kronecker product,  and $ s $ is the least common multiple of $ n $ and $ p $. $ \ltimes_{i=1}^n A_i \deq A_1\ltimes A_2\ltimes \cdots\ltimes A_n $.
\end{definition}
\begin{remark}
	All fundamental properties of the standard matrix product remain valid under STP \cite{cheng2010linear}. For notational simplicity, the symbol $\ltimes$ is omitted in the remainder.
\end{remark}

To get a multi-linear form of a Boolean function based on STP, we identify Boolean values by $ 1 \sim \delta_2^1 $ and $ 0 \sim \delta_2^2 $. 
\begin{lemma} \cite{cheng2010linear} \label{lemma: structure matrix}
	Any Boolean function $ f(x_1, x_2, \cdots, x_n): \Delta^n \rightarrow \Delta $ can be expressed in a multi-linear form as 
	\begin{equation}
	f(x_1, x_2, \cdots, x_n) = M_f x_1 x_2 \cdots x_n,
	\end{equation}
	where $ M_f \in  \mathcal{L}_{2\times 2^n}$ is a unique logical matrix, called the \textit{structure matrix} of $ f $.
\end{lemma}

A general SBCN with $ n $ state variables, $ m $ control inputs, and $ z $ subsystems, can be described as 
\begin{equation} \label{eq: sbcn}
\begin{cases}
x_1(t+1) = f_1^{\sigma(t)}(x_1(t), \cdots, x_n(t), u_1(t), \cdots, u_m(t))\\
\vdots \\
x_n(t+1) = f_n^{\sigma(t)}(x_1(t), \cdots, x_n(t), u_1(t), \cdots, u_m(t)),
\end{cases}
\end{equation}
where $ x_i(t) \in \Delta, u_j(t) \in \Delta $ denote states and control inputs respectively, and $ f_i^l : \Delta^{m+n} \rightarrow \Delta $ is the Boolean function associated with the state variable $ x_i $ in the $ l^{\textrm{th}} $ subsystem, $ 1 \le i \le n, 1 \le j \le m, 1\le l \le z $, while $ \sigma:  \bbN \rightarrow \Lambda = \{1, 2, \cdots, z \}$ is the switching law. 

Set $ x(t) \deq \ltimes_{i=1}^n x_i(t)$ and $ u(t) \deq \ltimes_{j=1}^m u_j(t) $. Note that  $ \ltimes_{i=1}^n: \Delta^n \rightarrow \Delta_{2^n}$ is a bijective mapping \cite{cheng2010linear}. Let $ N \deq 2^n $ and $ M \deq 2^m $, and we have $ x(t) \in \Delta_N $, $ u(t) \in \Delta_M $. 

The ASSR of the SBCN in \eqref{eq: sbcn} is given by,
\begin{equation} \label{eq: ASSR-SBCN}
    x(t+1) = L_{\sigma(t)}u(t)x(t),
\end{equation}
where $ L_{l} \in \mathcal{L}_{N \times MN}, 1 \le l \le z$, named the \textit{network transition matrix}, is computed by $ \text{Col}_j(L) = \ltimes_{i=1}^{n} \text{Col}_j(M_{f_i^{l}}), 1 \le j \le MN$, where $M_{f_i^{l}} \in  \mathcal{L}_{2\times MN}$ denotes the structure matrix of $ f_i^{l} $ in \eqref{eq: sbcn}. We refer readers to \cite{cheng2010linear,li2013minimum} for more details on how to compute the ASSR. To be consistent, we also identify the switching signal with a vector as $ l \sim \delta_z^l, l \in \Lambda $. We thus have equivalently $ \sigma: \bbN \rightarrow \Delta_z $.  Obviously, a non-switching BCN is just a special SBCN composed of a single sub-system.

\subsection{Graph, Path, and Cycle} \label{sec: graph def}
We introduce some fundamental concepts of graph theory in this part, mainly following the convention in \cite{cormen2009introduction}. 

A graph $ G $ is represented by an ordered pair $ (V, E) $, where $ V $ is a set of vertices, and $ E $ is a set of edges. A directed graph is graph with directed edges, and each edge $ e \in E $ from vertex $ v_i \in V$ to vertex $ v_j \in V$ is denoted by an ordered pair $ (v_i, v_j) $. Given an edge $ (v_i, v_j) $, $ v_i $ is called a \textit{predecessor} of $ v_j $, and $ v_j $ is a \textit{successor} of $ v_i $.
Additionally, each edge can be assigned a weight by a function $ w: E \rightarrow \bbR $. Denote the weight of an edge $ (v_i, v_j) $ by $ w(v_i, v_j) $.

\begin{definition} \label{def: path and cycle}
	We give the following definitions regarding paths and cycles on a weighted directed graph $ G = (V, E, w) $.
	\begin{itemize}
		\item A \textit{path }from vertex $ v_0 $ to vertex $ v_k $ is a sequence of vertices connected by edges, denoted by $ p = \left<v_0, v_1,  \cdots, v_k\right>, k \ge  0 $, where $ (v_i, v_{i + 1}) \in E $, $ 0 \le i < k $. Specially, if $ k = 0 $, $ p $ is an empty path. A \textit{simple path} is a path with no repeated vertices. An empty path is always simple. Let $ \psi(p) $ and $ |p| $ denote the number of edges and the number of  vertices in $ p $ respectively.
		\item A \textit{cycle} is a path whose first vertex and last vertex are the same, denoted by $ c = \left<v_0, v_1,  \cdots, v_k, v_0 \right>, k \ge 0$. A \textit{simple cycle} is a cycle which does not have any other repeated vertices except the first and last vertices, i.e., $ v_i \ne v_j, \forall 0 \le i, j \le k $ if $ i \ne j $.
		\item The weight of a path (or a cycle) $ p = \left<v_0, v_1,  \cdots, v_k\right>, k \ge 0 $, denoted by $ w(p) $, is the sum of weights of its constituent edges, given by
		\begin{equation} \label{eq: path weight}
			w(p) = \sum_{i=0}^{k-1} w(v_i, v_{i+1}).
		\end{equation}
	\end{itemize}
\end{definition}

\section{Problem Formulation} \label{sec: problem}
Like traditional control systems, constraints are common in BCNs. In therapeutics, we must avoid dangerous states of a GRN when applying radiation treatments in therapeutic practice. For example, the activated state of gene Wnt5a is undesirable because it can induce a melanoma metastasis \cite{pal2006optimal}. Various constraints have been considered in studies on BCNs \cite{li2015controllability,li2013minimum,fornasini2013optimal,li2014optimal}, which can be classified into three types: state constraints, input constraints, and transition constraints \cite{zhang2017finite} (named s\textit{tate-dependent input constraints} in \cite{yang2018stabilization}). We consider all these general constraints as well as switching constraints in the problem formulation as follows.

\begin{problem} \label{prob: 1}
	The IHOC of SBCNs \eqref{eq: ASSR-SBCN} with average cost subject to specific constraints is to solve the following constrained optimization problem:
	\begin{align} \label{eq: problem}
	\min_{\bu, \bsg} J(\bu, \bsg) = \lim_{T\rightarrow\infty} \frac{1}{T} \sum_{t=0}^{T-1} g(x(t), u(t), \sigma(t)), \nonumber\\
	\textrm{s.t.} \begin{cases}
	x(t+1) = L_{\sigma(t)}u(t)x(t) \\
	x(t) \in C_x \\
	u(t) \in C_u(x(t)) \\
	\sigma(t) \in C_{\sigma}(x(t)) \\
	x(0) = x_0 
	\end{cases},
	\end{align}
	where $ \bu = \big(u(t) \in \Delta_M\big)_{t=0}^{T-1} $ and $ \bsg = \big( \sigma(t) \in \Dz \big)_{t=0}^{T-1}  $ denote a control input sequence and a switching signal sequence respectively; $ g:  \DN \times \DM \times \Dz \rightarrow \bbR$ is the stage-wise cost function, which is assumed to be bounded;  $ C_x \subseteq \DN, C_u(x(t)) \subseteq \DM, C_{\sigma}(x(t)) \subseteq \Dz $ denote the state constraints, the state-dependent control input constraints, and the state-dependent switching signal constraints respectively; and  $ x_0 \in C_x$ is the initial state of the SBCN. 
\end{problem}

\rv{Most genetic diseases, like cancer, are caused by the dysfunction of certain genes or the abnormal interaction among genes. In a biomedical context, the optimal control of BCNs studied here thus corresponds to the systematic and effective external intervention of GRNs such that malignant cellular states, like metastasis, can be avoided, and healthy states can be retained \cite{pal2006optimal}. For example, the repression of the gene Wnt-5a could help to prevent the generation of a metastatic phenotype \cite{datta2007control}. Specifically, in the above stage cost function $ g(x, u, \sigma) $, the control input $ u $ may represent clinical treatment like drugs or exposure to certain radiations, and the switching $ \sigma $ may be applied by gene engineering tools like \textit{gene editing} or \textit{function perturbation} \cite{qian2008effect}. The design of the stage cost function is largely application-specific, which should essentially capture the benefits and costs of the intervention process \cite{pal2006optimal}, for example, the efficacy of a drug, the expense of medical treatment, and the potential side effects of such intervention. Since this paper focuses primarily on theoretical development, we choose the cost function arbitrarily for demonstration purposes.
}

\begin{remark} \label{rmk: constraints}
	Most existing studies, such as \cite{li2015controllability, li2013minimum} on BCNs and \cite{li2014optimal} on SBCNs, only consider input or switching signal constraints that are independent of states, for example, $ u(t) \in \bar{C}_u \subseteq \DM$, where $ \bar{C}_u $ is a constant set. Such state-independent constraints can be viewed technically as a special case of the general constraints in  \eqref{eq: problem}, e.g., $ C_u(x(t)) = \bar{C}_u, \forall x(t) \in C_x $. 
\end{remark}

\begin{remark}
	The boundedness of the per-stage cost $ g:  \DN \times \DM \times \Dz \rightarrow \bbR$ is commonly assumed, either explicitly or implicitly, in optimal control of BCNs and SBCNs like \cite{pal2006optimal, zhao2010optimal, li2014optimal,wu2019optimal}. Essentially, $ g $ has a finite number of possible inputs, and we can always assign sufficiently high (or low) but bounded costs to individual inputs \cite{fornasini2013optimal}. If we want to prohibit a state, an input, or a combination of the two completely, we can set hard constraints instead in Problem \ref{prob: 1}. 
\end{remark}

\section{Optimal State Transition Graph (OSTG)} \label{sec: ostg} 

\subsection{Construction of the OSTG} \label{sec: construction OSTG}
In Problem \ref{prob: 1}, we only care about the states that can be reached from $ x_0 $. \rv{Denote the set of states reachable from a state $ x \in \DN$ in exactly $ d$ steps by $ \cR(x, d) \subseteq \DN $.}  The full reachable set of $ x_0$ is computed by $\cR(x_0) = \cup_{d=0}^{N-1} \cR(x_0, d) $ because there are at most $ N $ reachable states in total \cite{zhao2010input,zhao2010optimal}. The reachability of BCNs (SBCNs) has been widely studied using algebraic approaches based on the ASSR, which is computationally expensive due to the calculation of matrix powers \cite{cheng2009controllability,zhao2010input,yu2018stabilizability,li2012reachability}. We will design a more efficient procedure based on the breadth-first search (BFS) \cite{cormen2009introduction} of a graph by iterative computations of $ \cR(\cdot, 1) $ (see Algorithm \ref{alg: OSTG}).  

If we view each admissible state as a vertex on a directed graph, then each edge $ (\dN^i, \dN^j) $ denotes a state transition from $ \dN^i $ to $ \dN^j $, and $ \cR(\dN^i, 1) $ comprises all successors of $ \dN^i $ in that graph. Such a graph is usually called a \textit{state transition graph} (STG) in the literature \cite{li2012reachability,cheng2014optimal,zhao2010input}. Consider Problem \ref{prob: 1}. Given two states $ \dN^i, \dN^j \in C_x $, $ \dN^j $ is reachable from $ \dN^i $ in one step if the following condition holds:
\begin{equation} \label{eq: reach algebraic}
\dN^j = L_{l}u\dN^i, \ \exists \dz^l \in C_{\sigma}(\dN^i), u \in C_u(\dN^i).
\end{equation}

Recall that both $ u \in \DM $ and $ x \in \DN $ are column vectors with a single entry being 1 and all others 0. A computationally economical way to get $ \cR(\dN^i, 1) $ following \eqref{eq: reach algebraic} is given below.

\begin{lemma} \label{lemma: successor}
	Consider Problem \ref{prob: 1}. Given a state $ \dN^i \in C_x$, its one-step reachable set is obtained according to \eqref{eq: reach algebraic} by
	\begin{align} \label{eq: successor}
	\cR(\dN^i, 1) = &\{  \textrm{Col}_i(\textrm{Blk}_k(L_l)) | \dM^k \in C_u(\dN^i),  \dz^l \in C_{\sigma}(\dN^i) \}  \nonumber\\
		& \cap C_x.
	\end{align}
\end{lemma}

The $ \cR(\cdot, 1) $ in Lemma \ref{lemma: successor} effectively gives the adjacency-list representation \cite{cormen2009introduction}  of the STG. Following the principle of BFS, the reachable set $ \cR(x_0) $ is obtained recursively:
\begin{equation} \label{eq: reachable set}
	\cR(x_0) = \{x_0\} \cup \cR(x_0, 1) \cup_{x' \in \cR(x_0, 1)} \cR(x').
\end{equation}

Next, we assign weights to these edges. For convenience, we call a pair of a control input and a switching signal, i.e., $ (\dM^k, \dz^l)$ in \eqref{eq: ad set}, an \textit{action}. Based on Lemma \ref{lemma: successor}, we collect all actions that can steer the SBCN \eqref{eq: ASSR-SBCN} from $ \dN^i $ to $ \dN^j $ into a set $ A^{ij} $, named the \textit{admissible action set}, given by
\begin{equation} \label{eq: ad set}
	A^{ij} = \{(\dM^k, \dz^l) |\textrm{Blk}_k(L_l)_{ji} = 1,  \dM^k \in C_u(\dN^i), \dz^l \in C_{\sigma}(\dN^i) \}.
\end{equation}
 
The optimal action is an action that enables the transition from $ \dN^i $ to $ \dN^j $ with the lowest cost:
\begin{equation} \label{eq: best choice}
	(\dM^{k^*}, \dz^{l^*}) = \argmin_{(\dM^k, \dz^l) \in A^{ij}} g(\dN^i, \dM^k, \dz^l).
\end{equation}

To ease statement, we abuse notations slightly in the remainder of the text by referring to a state of SBCN \eqref{eq: ASSR-SBCN} and the vertex bound to it in the STG both as $ \dN^i $. 

\begin{definition}\label{def: OSTG}
	Consider Problem \ref{prob: 1}. The OSTG with respect to the initial state $ x_0 $ is a directed weighted graph $ \Go = (V, E, w, x_0) $, where the vertex set is $ V = \cR(x_0) $, and the edge set is
	\begin{align} \label{eq: edges of OSTG}
		E = \{ (\dN^i, \dN^j) | \dN^i \in V,  \dN^j \in \cR(\dN^i, 1) \}.
	\end{align}
	Each edge $ (\dN^i, \dN^j) \in E $ has an action $ (\dM^{k^*}, \dz^{l^*}) $ yielded by \eqref{eq: best choice} and is assigned a weight by
	\begin{equation} \label{eq: OSTG weight}
		w(\dN^i, \dN^j) = g(\dN^i, \dM^{k^*}, \dz^{l^*}).
	\end{equation}
\end{definition}

\begin{remark}
	The OSTG is mainly inspired by the switching-input-state transfer graph \cite{li2012reachability} and the optimal input-state transfer graph \cite{zhu2018optimal} (both in a matrix form). In \cite{li2012reachability}, each vertex of the matrix-form graph is a triple $ (l, x, u), l\in \Lambda, x \in \DN, u \in \DM $, resulting in $ zMN $ vertices in total. By contrast, our OSTG focuses only on states reachable from $ x_0 $ and has at most $ N $ vertices in an adjacency-list representation \cite{cormen2009introduction}.
\end{remark}

Based on \eqref{eq: successor}, we detail the construction of the OSTG in Algorithm \ref{alg: OSTG}, whose skeleton is based on the BFS of a graph using a first-in-first-out (FIFO) queue \cite{cormen2009introduction}. \rv{BFS starts from a given vertex ($ x_0 $ in Line \ref{line: init}) and explores all neighbor vertices (i.e., \textit{successor}s) of this vertex before proceeding to further vertices (see the loop: Line \ref{line: for} - \ref{line: forend}). Once a vertex is visited, it is marked and will never be visited again (Line \ref{line: if1} - \ref{line: ifend1}). Note that, in Algorithm \ref{alg: OSTG}, the neighbors of a vertex are computed on the fly. To retain the breadth-first semantic, a FIFO queue is used: when a vertex is being visited, all its unvisited neighbors are pushed into the queue at the tail (\textit{enqueue}), and the next vertex to be visited is queue head, which is accessed and removed from the queue (\textit{dequeue}). Clearly, the set of neighbors of a vertex $ \dN^i $ is just $ \cR(\dN^i, 1) $ in \eqref{eq: successor}. The FIFO queue operation ensures the vertices are visited in a layered manner according to their distances to the root vertex $ x_0 $. At the end, we will get the neighbors for all vertices reachable from $ x_0 $, i.e.,  $ \cR(\dN^i, 1), \forall \dN^i \in \cR(x_0) $, which essentially forms the adjacency-list representation \cite{cormen2009introduction} of the OSTG in Definition \ref{def: OSTG}. }

\begin{algorithm}[htb]
	\caption{Construction of the OSTG for Problem \ref{prob: 1}} \label{alg: OSTG}
	\begin{algorithmic}[1] 
		\Input Problem \ref{prob: 1}: the SBCN \eqref{eq: ASSR-SBCN} and the constraints \eqref{eq: problem}
		\Output Adjacency-list representation of the OSTG
		\State Initialize a FIFO queue $ Q \gets \{ x_0\}$ \label{line: init}
		\State Initialize a Boolean array $ B $ of size $ N+1 $ with FALSE\footnotemark
		\State Create a dictionary $ D: \bbN\times\bbN \rightarrow \emptyset $  
		\While{$ Q $ is not empty} \label{line: while}  
			\State $ \dN^i \gets $ Dequeue($ Q $), \  $ \cR(\dN^i, 1) \gets \emptyset $ \label{Line: dequeue}
			\ForAll{$ \dM^k \in C_u(\dN^i), \dz^l \in C_{\sigma}(\dN^i) $} \Comment{\rv{see  \eqref{eq: successor}}} \label{line: for}
				\State $ \dN^j \gets \textrm{Col}_i(\textrm{Blk}_k(L_l)) $ \label{line: j}
				\If{$ \dN^j \in C_x $}
					\State $ \cR(\dN^i, 1) \gets \cR(\dN^i, 1) \cup \{  \dN^j  \}$  
					\State $ D[i, j] \gets D[i, j] \cup \{(\dM^k, \dz^l) \}$
					\If{$ B[j] = $ FALSE} \Comment{\rv{mark $ B[j] $}}\label{line: if1}
						\State $ B[j] \gets  $ TRUE, \  Enqueue($ Q, \dN^j $)  \label{line: enqueue}
					\EndIf \label{line: ifend1}
				\EndIf
			\EndFor \label{line: forend}
		\EndWhile \label{line: whileend}
		\State Compute the optimal action and the minimum weight of each edge with $ A^{ij} \deq D[i, j] $ according to \eqref{eq: best choice} and \eqref{eq: OSTG weight}  \label{line: edge info}
	\end{algorithmic}
\end{algorithm}
\footnotetext{Arrays in all algorithms of this paper start indexing from 0.}

\textit{Time Complexity Analysis:} In Algorithm \ref{alg: OSTG}, the \textbf{while} loop (Line \ref{line: while}--\ref{line: whileend}) executes $ |V| $ times, and the inner \textbf{for} loop (Line \ref{line: for}--\ref{line: forend}) runs no more than $ zM $ times, \rv{since each state transits to at most $ zM $ succeeding states in one step}. Finally, Line \ref{line: edge info} computes the stage cost $ g $ for at most $ zM|V| $ transitions to solve \eqref{eq: best choice} and \eqref{eq: OSTG weight}. The time complexity of Algorithm \ref{alg: OSTG} is thus $ O(zM|V|) $, or equivalently, $ O(zMN) $, since $ |V| \le N $.

\subsection{An Illustrative Example} \label{sec: example}
We use an SBCN adapted from \cite{li2014optimal} to illustrate the OSTG.

\begin{example} \label{example: 1}
	Consider the following SBCN with $ n=3 $ states, $ m=1 $ control input, and $ z=2 $ subnetworks:
	\begin{equation} \label{eq: example 1 model}
		x_i = f_i^{\sigma(t)}(x_1(t), x_2(t), x_3(t), u(t)), \  i =1, 2, 3
	\end{equation}
	where $ \sigma: \bbN \rightarrow \{1, 2\} $ is the switching signal, and
	\begin{align*}
		&f_1^1 \deq (u \oplus x_1) \land (x_2 \leftrightarrow x_3),  &f_1^2 &\deq x_1 \lor (x_2 \to x_3), \\
		&f_2^1 \deq \lnot x_3,  &f_2^2 &\deq \lnot x_3, \\
		&f_3^1 \deq (u \oplus x_1) \lor (x_2 \land x_3),  &f_3^2 &\deq (u \oplus x_1) \land (x_2 \lor x_3).
	\end{align*}
	Suppose that the constrains are given by
	\begin{equation} \label{eq: constraints ex1}
		\begin{cases}
			C_x = \delta_8 \{ 1\; 2\; 3\; 5\; 6\; 7\; 8  \} \\
			C_u(x) = \Delta, \forall x \in C_x, \\
			C_{\sigma}(x) = \begin{cases}
			\{1 \}, &\forall x \in \delta_8[1\; 2\; 5], \\
			\{1, 2\}, &\forall x \in \delta_8[3 \; 6\; 7 \;8]
			\end{cases}
		\end{cases}
	\end{equation}
	i.e., the state $ \delta_8^4 $ should be avoided, and only the first subnetwork can be activated for states $\delta_{8}^1,  \delta_{8}^2 $ and $ \delta_{8}^5 $, while there are no constraints on the control input. 
	We adopt an arbitrary stage cost function for illustration purpose:
	$ g(x(t), u(t), \sigma(t)) = x(t) Q_x x(t) + u(t)Q_uu(t) + \sigma(t)Q_\sigma \sigma(t), $
	where $ Q_x = \text{diag}(5, 3, 4, 0, 1, 3, 0, 1), Q_u = \text{diag}(3, 1) $, and $ Q_{\sigma} = \text{diag}(1, 2) $ are diagonal cost matrices.
\end{example}

We first get the ASSR for \eqref{eq: example 1 model} in form of \eqref{eq: ASSR-SBCN} as $ L_1 = \delta_8 [7\; 6\; 8\; 6\; 3\; 5\; 7\; 1\; 3\; 5\; 7\; 1\; 7\; 6\; 8\; 6] $ and $ L_2 = \delta_8 [4\; 2\; 4\; 2\; 3\; 5\; 3\; 2\; 3\; 1\; 3\; 2\; 4\; 6\; 4\; 2]. $
\rv{We illustrate the execution of one \textbf{while} loop in Algorithm \ref{alg: OSTG} with the initial state $ x_0 = \delta_8^1 $ as follows. The FIFO queue is initialized as $ Q = \{\delta_8^1\} $. In Line \ref{Line: dequeue}, the dequeue operation yields $ \dN^i \gets  \delta_8^1$. 
Its succeeding states under constraints \eqref{eq: constraints ex1} can be obtained by \eqref{eq: successor} with $ i =1, k \in \{1, 2\}, l \in \{1\} $. For example, given $ i = 1, k =1 $ and $ l = 1 $, we get one successor $  \textrm{Col}_1(\textrm{Blk}_1(L_1)) = \delta_8^7$ (i.e., $ \dN^j \gets \delta_{8}^7 $ in Line \ref{line: j}). Since $ \delta_{8}^7 $ is not visited yet (i.e., $ B[7] =  $ FALSE in Line \ref{line: if1}), it is pushed  into $ Q $ (Line \ref{line: enqueue}). This procedure is repeated for $ i =1, k =2,  $ and $ l =1 $ to get another successor $  \textrm{Col}_1(\textrm{Blk}_2(L_1)) = \delta_8^3 $. We then have $ \cR(\delta_8^1, 1) = \{ \delta_8^7, \delta_8^3 \}, $, and the queue is now $ Q = \{ \delta_8^7, \delta_8^3\} $.}

\begin{figure}[tb]
	\centering
	\includegraphics[width=45mm]{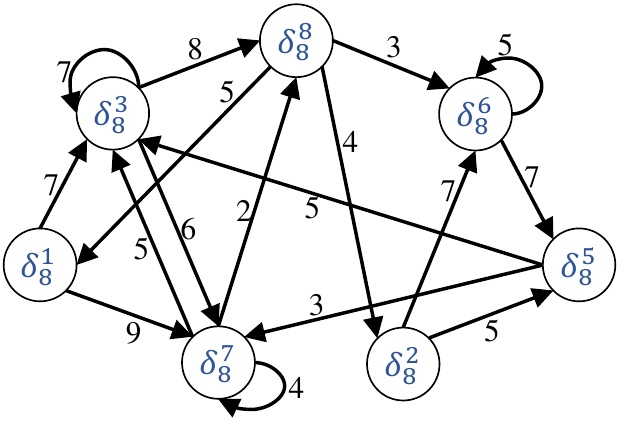} 
	\caption{ The OSTG in Example \ref{example: 1} with the initial state $ x_0 = \delta_8^1 $. Each edge is annotated with its weight according to \eqref{eq: OSTG weight}.}
	\label{fig: OSTG of example 1}
\end{figure}

Applying Algorithm \ref{alg: OSTG}, we build the OSTG for Example \ref{example: 1}, illustrated in Fig. \ref{fig: OSTG of example 1}, and compute its weights by \eqref{eq: OSTG weight}. The optimal action \eqref{eq: best choice} for each edge is not shown in Fig. \ref{fig: OSTG of example 1} for clarity purpose. An example is $ A^{65} = \{(\delta_2^1, \delta_2^1), (\delta_2^1, \delta_2^2)\}  $, and the optimal action to transit from $ \delta_{8}^6 $ to $ \delta_{8}^5 $ is $ (\delta_2^1, \delta_2^2) $ with a minimum cost of 2. Besides, though we have $ \cR(x_0) = C_x $ in Fig. \ref{fig: OSTG of example 1}, i.e., all admissible states can be reached from the initial state $ x_0 = \delta_8^1 $, it is typically not true for large networks.

\section{Solve IHOC with Average Cost using the OSTG} \label{sec: algrithms}
It is first shown in \cite{zhao2010optimal} that the state trajectory of a BCN under IHOC will converge to a cycle in the input-state space. We adopt a similar idea, but we prove the connection between Problem \ref{prob: 1} and an optimal cycle (see Definition \ref{def: MMC}) in the OSTG rigorously. More importantly, we propose a novel method based on a minimum-mean cycle algorithm in graph theory to locate the optimal cycle and to obtain the optimal solution to Problem \ref{prob: 1} via state feedback with exceptional efficiency. 

\subsection{Path Decomposition}
To handle the infinitely long state trajectory encountered in Problem \ref{prob: 1}, we first give the following proposition.

\begin{proposition} \label{prop: decomposition}
	Consider a directed graph $ G = (V,E) $. Given any non-simple path $ p$ from $ v_0 \in V $ to $ v_k \in V$ in $ G $, $ p $ can be decomposed into a list of simple cycles, $ c_1, c_2, \cdots, c_q, q \ge 1, $ and a simple path $ \ps $ from $ v_0 $ to $ v_k $, such that 
	\begin{align} \label{eq: decomposition}
	 	\cE(p) =  \left(\biguplus_{i=1}^q \cE(c_i)\right) \uplus \cE(\ps),
	\end{align}
	where $ \cE(p) $ denotes the set of edges (including repeated ones)\footnote{Also known as a \textit{multiset} in mathematics.} of the path $ p $, and $ \uplus $ denotes the union operation preserving duplications.
\end{proposition}
\begin{IEEEproof}
	Set $ p^{(1)} \deq p=  \left<v_0, v_1,  \cdots, v_k\right>, k > 0 $ for notational simplicity.  Since $ p^{(1)} $ is a non-simple path, it must contain duplicate vertices. Assume a pair of such repetitive vertices is $ v_i $ and $ v_j$  with $0 \le i < j  \le k $ that satisfies $ v_{i'}  \ne v_{j'}, \forall i < i', j' < j$. Note that such $ v_i $ and $ v_j $ always exist because if $ v_{i'}  = v_{j'}  $, we can let $ i=i' , j = j'$ and repeat. 
	We decompose $ p^{(1)} $ into a simple cycle $ c^{(1)} = \left<v_i, v_{i+1}, \cdots, v_j \right> $ and a remainder path $ p^{(2)} = \left<v_0, v_1, \cdots, v_{i-1}, v_j, v_{j+1}, \cdots, v_k \right>$. Note that we leave the \textit{last} vertex of the cycle, i.e., $ v_j $, in the remainder $ p^{(2)} $ to form a path from $ v_0 $ to $ v_k $ unless $ p^{(1)} $ itself is a simple cycle, in which case $ p^{(2)} $ is empty. It is easy to see that $ \cE(p^{(1)})  = \cE(c^{(1)}) \uplus \cE(p^{(2)})$. Specially, even $ c^{(1)} $ begins with $ v_0 $, a non-empty $ p^{(2)} $ is still a path in $ G $ from $ v_0 $ to $ v_k $, since we leave the last vertex, i.e., $ v_0 $ here, in $ p^{(2)} $.  The same reasoning applies if $ c^{(1)} $ ends with $ v_0 $.
	
	Similarly, if the remainder path $ p^{(2)} $ is still non-simple, we can apply the above decomposition procedure to $ p^{(2)} $ again and get $ \cE(p^{(2)})  = \cE(c^{(2)}) \uplus \cE(p^{(3)})$. This process will be repeated for $ q $ times until $ p^{(q+1)} $ is a simple path. Note that $ q $ must be finite, because our operation guarantees $ |p^{(l + 1)}| < |p^{(l)} |, \forall l \ge 1$. It follows obviously that $ \cE(p^{(1)})  = \cE(c^{(1)}) \uplus \cE(p^{(2)}) = \cE(c^{(1)}) \uplus  \cE(c^{(2)}) \uplus \cE(p^{(3)})  = \cE(c^{(1)}) \uplus \cdots \uplus \cE(c^{(q)}) \uplus \cE(p^{(q+1)})$.  In the $ l $-th operation, $\forall 1 \le l \le q $, we end with $ p^{(l+1)} $, a simple path from $ v_0 $ to $ v_k $, which can be empty.
	
	Recall that $ p^{(1)} \deq p $. Simply set $ c_i \deq c^{(i)}, \forall 1 \le i \le q$ and $ \ps \deq  p^{(q+1)}$. Then we can get \eqref{eq: decomposition}. 
\end{IEEEproof}

\begin{remark} \label{rmk: nonunique decomposition}
	The path decomposition in Proposition \ref{prop: decomposition} may not be unique. Nevertheless, Eq. \eqref{eq: decomposition} always holds for any qualified decomposition. 
\end{remark}

\begin{corollary} \label{coro: weight decomposition}
	Following Proposition \ref{prop: decomposition}, since the edge weights are fixed, Eq. \eqref{eq: decomposition} implies
	\begin{equation} \label{eq: decompostion weight}
		w(p) = \sum_{i=1}^{q} w(c_i) + w(\ps).
	\end{equation}
\end{corollary}

We use an example to explain the above path decomposition.
\begin{example} \label{example: 2}
	Recall Example \ref{example: 1} and the OSTG in Fig. \ref{fig: OSTG of example 1}. Consider a path $ p $ starting from $ x_0 $ and composed of 14 vertices: $ p = \left<\delta_{8}^1 , \delta_{8}^3 , \delta_{8}^8 , \delta_{8}^6 , \delta_{8}^5 , \delta_{8}^7 , \delta_{8}^8 ,\delta_{8}^2  ,\delta_{8}^5 ,\delta_{8}^7 ,\delta_{8}^7 ,\delta_{8}^3, \delta_{8}^8, \delta_{8}^2 \right> $.
	We can extract three simple cycles from $ p $: $ c_1 = \left< \delta_8^8, \delta_8^6 , \delta_8^5 , \delta_8^7 , \delta_8^8  \right> $, $ c_2 = \left< \delta_{8}^7, \delta_8^7  \right> $, and $ c_3 = \left< \delta_8^3 , \delta_8^8, \delta_8^2 , \delta_8^5, \delta_8^7, \delta_{8}^3 \right> $. The remainder path is  $ \ps= \left< \delta_{8}^1, \delta_{8}^3, \delta_{8}^8, \delta_{8}^2 \right> $. It is easy to verify \eqref{eq: decomposition} and \eqref{eq: decompostion weight}. \rv{Note that the path decomposition is not unique (see Remark \ref{rmk: nonunique decomposition}). Nevertheless, the exemplified extraction of simple cycles in a left-to-right manner facilitates programming implementation.}
\end{example}

\subsection{Solution based on Minimum-Mean Cycle (MMC)}

We first give a proposition relating a state trajectory of the SBCN to a path in its OSTG.

\begin{proposition} \label{prop: trajectory->path}
	Consider Problem \ref{prob: 1} and its OSTG $ \Go=(V, E, w, x_0) $. Given any state trajectory of the SBCN  $ s =(x(0), x(1), \cdots, x(T)) = \big( \dN^{i_t} \big)_{t=0}^{T}, \dN^{i_0} = x_0$, steered by an action sequence $ \bm{a} = \big(  (u(t), \sigma(t)) \big)_{t=0}^{T-1} $, then $ p= \big<\dN^{i_t} \big>_{t=0}^{T} $ is a path in $ \Go$, which satisfies $ w(p) \le Q(s, \bm{a}) $, where 
	\begin{equation} \label{eq: total cost}
		Q(s, \bm{a}) = \sum_{t=0}^{T-1} g(x(t), u(t), \sigma(t)).
	\end{equation}
	$ w(p) = Q(s, \bm{a}) $ holds if each action $ (u(t), \sigma(t)) $ is the optimal one in \eqref{eq: best choice} from state $ \dN^{i_t} $ to state $ \dN^{i_{t+1}}$, $ \forall 0 \le t \le T-1 $. 
\end{proposition}
\begin{IEEEproof}
	Since $ s $ is a state trajectory starting from $ \dN^{i_0} = x_0 $, we have $ \dN^{i_t} \in \cR(x_0), \forall 0 \le t \le T $ and $ \dN^{i_{t+1}} \in \cR(\dN^{i_{t}}, 1), \forall 0 \le t < T  $. By Definition \ref{def: OSTG} of the OSTG, $ p$ must be a path in $ \Go $. Recall the optimality of the OSTG implied by \eqref{eq: best choice} and \eqref{eq: OSTG weight}. For any state transition from $ \dN^{i_t}$ to $ \dN^{i_{t+1}}, 0 \le t < T $, driven by $ (u(t), \sigma(t)) $, we have $w(\dN^{i_t}, \dN^{i_{t+1}}) \le g(\dN^{i_t}, u(t), \sigma(t))$, whose equality holds with the optimal action defined in \eqref{eq: best choice}.  It follows directly from \eqref{eq: path weight} and \eqref{eq: total cost} that $ w(p) \le Q(s, \bm{a}) $, and the equality is true if each action is the optimal one in \eqref{eq: best choice}.
\end{IEEEproof}

A minimum-mean cycle in a graph is defined as follows.

\begin{definition} \label{def: MMC}
	Given a directed weighted graph $ G = (V, E, w) $ and a path (possibly a cycle) $ p $ in $ G $, \rv{denote the \textit{average weight} of $ p $ by $ \bar{w}(p) = \frac{w(p)} {\psi(p)} $}. Let the set of cycles in $ G $ be $ \cC(G) $.  A cycle $ c^* $ is called the \textit{minimum-mean cycle} (MMC)  in $ G $ if it satisfies the following condition:
	\begin{equation} \label{eq: MMC}
		\bw(c^*) = \min_{c \in \cC(G)} \bw(c).
	\end{equation}
\end{definition}

\begin{lemma} \label{lemma: simple MMC}
	Following Definition \ref{def: MMC}, if an MMC exists (i.e.,  $ \cC(G) \ne \emptyset$), there must exist a simple cycle which is an MMC.
\end{lemma}
\begin{IEEEproof}
	Suppose $ c^* $ is an MMC. If $ c^* $ is not simple, Proposition \ref{prop: decomposition} tells that we can decomposes $ c^* $ into $ c^* = \biguplus_{i=1}^q c_i, q \ge 2 $, where $ c_i, 1\le i \le q, $ are all simple cycles. From \eqref{eq: decomposition} and \eqref{eq: decompostion weight}, we have
	\begin{equation}
		\psi(c^*) = \sum_{i=1}^{q} \psi(c_i), \quad w(c^*) = \sum_{i=1}^{q} w(c_i),
	\end{equation}
	which further leads to
	\begin{equation}
		\bw(c^*) = \frac{\psi(c_1)\bw(c_1) +\psi(c_2)\bw(c_2) + \cdots + \psi(c_q)\bw(c_q)}{\psi(c_1) + \psi(c_2) + \cdots + \psi(c_q)},
	\end{equation}
	which is a convex combination of $ \bw(c_1), \bw(c_2), \cdots,  $ and $ \bw(c_q) $. Hence, there holds $ \bw(c^*) \ge \min_{i=1}^q \bw(c_i) $. From \eqref{eq: MMC}, it implies that $ \bw(c^*) = \bw(c_j), j = \argmin_{i=1}^q \bw(c_i)  $; that is, the simple cycle $ c_j$ is also an MMC by Definition \ref{def: MMC}.
\end{IEEEproof}

\begin{remark} \label{rmk: MMC}
	A conclusion similar to Lemma \ref{lemma: simple MMC} in the input-state space is proved in \cite[Proposition 4.4]{zhao2010optimal}. Here we prove its correctness in a graph with a different method, i.e.,  through path decomposition and convex combination. 
\end{remark}

\rv{Note that, since our problem setting assumes that infinite state trajectories are applicable, which corresponds to infinitely long paths in the OSTG $ \Go $, and all stage costs are bounded, $ \Go $ must have cycles and thereby an MMC.  Lemma \ref{lemma: simple MMC} further tells that there must exist a simple MMC in $ \Go $.}

\rv{
\begin{example}\label{ex: MMC}
	Consider the OSTG shown in Fig. \ref{fig: OSTG of example 1}, which has multiple cycles. We list three examples of simple cycles as follows: $ c_1 = \delta_{8} \left< 7, 8, 2, 5, 7 \right>, c_2 = \delta_{8} \left< 3, 3 \right> $, and $ c_3 = \delta_{8} \left< 3, 8, 6, 5, 3 \right> $. It is easy to calculate that $ \bw(c_1) = 3.5, \bw(c_2) = 7 $, and $ \bw(c_3) = 5.75 $. As we will show in Example \ref{ex: solve toy problem}, $ c_1 $ is actually an MMC of the OSTG.
\end{example}
}

\begin{definition} \label{def: convergence}
	Consider Problem \ref{prob: 1} and its OSTG $ \Go $. We say that the state trajectory $ s $ of SBCN \eqref{eq: ASSR-SBCN} converges to a cycle $ c^* $ in $ \Go $ if the path in $ \Go $ that corresponds to $ s $ keeps repeating $ c^* $ after a finite time $ T_{c^*} < \infty$. 
\end{definition}

\begin{theorem} \label{thm: optimal}
	Consider Problem \ref{prob: 1} and its OSTG $ \Go = (V, E, w, x_0) $. If $ c^* $ is a simple MMC in $ \Go $, and $ J^* $ is the minimum objective value of Problem \ref{prob: 1}, then an action sequence $\ba^* =  (\bu^*, \bsg^*) $ is a minimizer to \eqref{eq: problem}, i.e., $ J^* = J(\bu^*, \bsg^*)$, if the state trajectory that it induces converges to $ c^* $, and each action $ (u(t), \sigma(t)), \forall t \ge 0, $ is the optimal one determined by \eqref{eq: best choice}. The optimal objective value is $ J^* =  \bw(c^*) $.
\end{theorem}

\begin{IEEEproof}
	Given any feasible action sequence $\ba =  (\bu, \bsg) $ to Problem \ref{prob: 1}, let $ s = ( \dN^{i_t} )_{t=0}^{T}$, $ \dN^{i_0} = x_0, $ be the resultant state trajectory of the SBCN starting from $ x_0 $. According to Proposition \ref{prop: trajectory->path}, consider the associated path $ p= \big<\dN^{i_t} \big>_{t=0}^{T} $ in $ \Go $: we have $ Q(s, \ba) \ge w(p) $. Next, we first show (i) $ J(\bu, \bsg)  \ge \bw(c^*)$, and then (ii) $J(\bu^*, \bsg^*)  = \bw(c^*)  $.
	
	(i) Since we target IHOC, $ p $ is a non-simple path. Corollary \ref{coro: weight decomposition} implies that there exists $ q \ge 1$ such that
		\begin{equation*} 
		\frac{w(p)}{T} = \frac{1 }{T} \sum_{i=0}^{q} w(c_i) + \frac{w(\ps)}{T} = \frac{1 }{T} \sum_{i=0}^{q} \psi(c_i) \bw(c_i) + \frac{w(\ps)}{T}, 
		\end{equation*}
		where $ \ps $ is a simple path, and each $ c_i $ is a simple cycle. Recall the definition of an MMC in \eqref{eq: MMC}, and we further have
		\begin{align}\label{eq: infty}
		\frac{w(p)}{T} & \ge \frac{1}{T} \sum_{i=0}^{q} \psi(c_i) \bw(c^*) + \frac{w(\ps)}{T}	\nonumber\\
			& = \frac{T - \psi(\ps)}{T} \bw(c^*) + \frac{w(\ps)}{T}.
		\end{align} 
		The simple path $ \ps $ satisfies $ |\ps| \le |V| $ and $ \psi(\ps) < |V| $, which ensures that $ w(\ps) $ is bounded. We have 
		\begin{align} \label{eq: optimality criterion}
		J(\bu, \bsg) &= \lim_{T\rightarrow\infty} \frac{Q(s, \ba)}{T} \ge \lim_{T\rightarrow\infty}  \frac{w(p)}{T}	\nonumber\\
		&\ge \lim_{T\rightarrow\infty} \frac{T - \psi(\ps)}{T} \bw(c^*) + \frac{w(\ps)}{T} \nonumber\\
		&= \bw(c^*),
		\end{align}
		\rv{where the first inequality follows from Proposition \ref{prop: trajectory->path}; the second inequality follows from \eqref{eq: infty}; and the last equality is a direct result of the limit operator.}
		
	(ii) Suppose the state trajectory induced by $\ba^* =  (\bu^*, \bsg^*) $ is $ s^* $ and the corresponding path in $ \Go $ is $ p^* $. Since $ s^* $ converges to $ c^* $, $ p^* $ must begin with a finite sub-path $ \pt $, after which $ p^* $ keeps repeating $ c^* $ by Definition \ref{def: convergence}. Therefore, we have
	\begin{equation}
		w(p^*) = w(\pt) + k w(c^*) + w(p'),
	\end{equation}
	where $ p' $ is a sub-path of $ c^* $ if $ s^* $ has not finished the last cycle, and the number of cycles that $ s^* $ has finished is
	\begin{equation}
		k = \frac{T - \psi(\pt) - \psi(p')}{\psi(c^*)}.
	\end{equation}
	Note that $ Q(s^*, \ba^*) = w(p^*)$ because each action is chosen as an optimal one in \eqref{eq: best choice}. We thus have
	\begin{align} \label{eq: J u* w*}
		J(\bu^*, \bsg^*) &= \lim_{T \rightarrow \infty} \frac{Q(s^*, \ba^*)}{T}  \nonumber\\
			&=\lim_{T \rightarrow \infty} \frac{w(\pt)}{T} + \frac{T - \psi(\pt) - \psi(p')}{T\psi(c^*)}w(c^*) + \frac{w(p')}{T}   \nonumber\\
			&= \frac{w(c^*)}{\psi(c^*)} \nonumber\\
			&= \bw(c^*)
	\end{align}
	Now we have finished the proof of the two claims, which state together that $ J(\bu, \bsg) \ge J(\bu^*, \bsg^*)  $ for any feasible solution $ (\bu, \bsg)  $ to Problem \ref{prob: 1}. Thus, we have $ J^* = J(\bu^*, \bsg^*) = \bw(c^*) $, which can be obtained with the action sequence $ \ba^* $.
\end{IEEEproof}

\begin{remark}
	Technically, the optimal state trajectory only needs to converge to an MMC, which is not necessarily a simple one, though Lemma \ref{lemma: simple MMC} guarantees the existence of a simple MMC. We require $ c^* $ to be a simple one in Theorem \ref{thm: optimal} mainly to facilitate subsequent state-feedback control law design.  Note that the optimal value $ J^* $ in Theorem \ref{thm: optimal} depends on $ x_0 $ \cite{zhao2010optimal,fornasini2013optimal}, since the OSTG $ G_0 $ depends on $ x_0 $ by $ V = \cR(x_0) $. Given a specific $ x_0 $, the optimal solution $ (\bu^*, \bsg^*) $ to Problem \ref{prob: 1} may not be unique, because the simple MMC $ c^* $ and the one-step optimal action \eqref{eq: best choice} can be nonunique.
\end{remark}

For notational simplicity, denote the best action \eqref{eq: best choice} associated with each edge $ (\dN^i, \dN^j) $ in the OSTG by $ (u_{ij}, \sigma_{ij}) $. Following Theorem \ref{thm: optimal}, we show that the optimal solution to Problem \ref{prob: 1} can be expressed by a static state-feedback law (called a \textit{stationary policy} in \cite{wu2019optimal}) in the theorem below.

\begin{theorem}\label{thm: state feedback}
	Consider Problem \ref{prob: 1} and its OSTG $ \Go = (V, E, w, x_0) $. The following two statements are true:
	\begin{enumerate}
		\item  A state trajectory $ s^*$ starting from $ x_0 = \dN^{i_0}$ exists that converges to a simple MMC $ c^* $ without previously entering any other cycles, that is, 
		\begin{equation} \label{eq: optimal trajectory}
			s^* = \dN(i_0^*, i_1^*, \cdots,  i_{\alpha-1}^*, i_{\alpha}^*, i_{\alpha+1}^*, \cdots, i_{\beta}^*,  i_{\alpha}^*, i_{\alpha+1}^*, \cdots ),
		\end{equation}
		where the transient path $ \pt = \dN\big< i_0^*, i_1^*, \cdots, i_{\alpha}^*\big>$ is simple; $ c^* = \dN\big< i_{\alpha}^*, i_{\alpha+1}^*, \cdots, i_{\beta}^*,   i_{\alpha}^* \big>,0 \le \alpha \le \beta < |V|, $ is a simple MMC; and $ i_0^* \deq i_0$. Besides, $ s^* $ satisfies that $ i_{t_1}^* \ne i_{t_2}^*, \forall t_1 \ne t_2, 0 \le t_1, t_2 \le \beta $.
		\item The state trajectory \eqref{eq: optimal trajectory} is optimal if it is driven by a state-feedback control and switching law as follows:
		\begin{equation} \label{eq: state feedback law}
			u^*(t) = K_u x(t), \ \sigma^*(t) = K_{\sigma} x(t),
		\end{equation}
		with $ K_u \in \cL_{M \times N}  $ and $ K_{\sigma} \in \cL_{z \times N} $. 
		Define $ i_{\beta+1}^* \deq  i_{\alpha}^* $. For $ 0 \le t \le \beta $, $ K_u $ and $ K_{\sigma} $ are constructed by 
		\begin{equation} \label{eq: state feedback gain}
			\textrm{Col}_{i_t^*}(K_u) = u_{i_t^* i_{t+1}^*}, \   \textrm{Col}_{i_t^*}(K_{\sigma}) = \sigma_{i_t^* i_{t+1}^*},
		\end{equation}
		and the other columns of $ K_u$ and $ K_{\sigma} $ are arbitrarily set.
	\end{enumerate}
\end{theorem}
\begin{IEEEproof}
	Lemma \ref{lemma: simple MMC} states that there must exist a simple MMC, say $ c^* $,  in $\Go $. Besides, $ c^* $ can be reached from $ x_0  $ because $ V = \cR(x_0) $. Suppose a path progressing from $ x_0 $ to the MMC $ c^* $ is $ p= \dN \big< i_0, i_1, \cdots, i_{\tau} \big>$, where $ \dN^{i_{\tau}} \in c^* $ is a vertex in $ c^* $, and no other vertex of $ p $ lies in $ c^* $. We can then construct $ s^* $  as follows.
	\begin{itemize}
		\item  If $ p $ is not simple, then by Proposition \ref{prop: decomposition} a simple path from $ \dN^{i_0} $ to $ \dN^{i_{\tau}} $ can be obtained from $ p$, denoted by $ \ps = \dN \big< i_0^*, i_1^*, \cdots, i_{\alpha-1}^*,  i_{\alpha}^* \big> $ with $ i_0^* \deq i_0 $ and $ i_{\alpha}^* \deq i_{\tau} $. 
		\item  Since $ \dN^{i_{\alpha}^*} = \dN^{ i_{\tau}} $ is a vertex in $ c^* $, the MMC $ c^* $ can be expressed
		as $ c^* = \dN\big< i_{\alpha}^*, i_{\alpha+1}^*, \cdots, i_{\beta}^*,   i_{\alpha}^* \big>, \alpha \le \beta $.
	\end{itemize}
	Now combing $ p_s $ and $ c^* $, we get $ s^* $ in \eqref{eq: optimal trajectory}. The transient path $ \pt $ in Theorem \ref{thm: state feedback} is just the simple path $ \ps $ here.
	Additionally, since both $ \ps $ and $ c^* $ are simple and no vertex in $ \ps $ belongs to $ c^* $ except $ \dN^{i_{\alpha}^*} $, we have $ i_{t_1}^* \ne i_{t_2}^*, \forall t_1 \ne t_2, 0 \le t_1, t_2 \le \beta $. Thus, the first statement is justified.

	To make $ s^* $ an optimal trajectory, Theorem \ref{thm: optimal} states that we just need to apply the optimal action \eqref{eq: best choice} for each transition of $ s^* $. The optimal control sequence to Problem \ref{prob: 1} is thus 
	\begin{align*}
		\bu^* = (&u_{i_0^* i_{1}^*}, u_{i_1^* i_{2}^*},  \cdots, u_{i_{\alpha}^* i_{\alpha +1}^*}, \cdots, u_{i_{\beta-1}^* i_{\beta}^*}, u_{i_{\beta}^* i_{\alpha}^*}, \nonumber\\
			 &u_{i_{\alpha}^* i_{\alpha +1}^*}, \cdots, u_{i_{\beta-1}^* i_{\beta}^*}, u_{i_{\beta}^* i_{\alpha}^*}, \cdots).
	\end{align*}
	Since $ p' \deq \dN\big<i_0^*, i_1^*, \cdots,i_{\alpha}^*,\cdots, i_{\beta}^* \big> $ is a simple path, we can define a function $ \kappa_u: \DN \rightarrow \DM $ to map $ \dN^{i_t^*} $ to $ u_{i_t^* i_{t+1}^*}$ for $ 0 \le t \le \beta $ with $ i_{\beta+1}^* \deq i_{\alpha}^* $. Note that there holds $ K_u \dN^{i_t^*} = \textrm{Col}_{i_t^*}(K_u)$. Hence, $ \kappa_u $ can be expressed by $\kappa_u(x(t)) = K_u x(t)$ for any state (vertex) $ x(t) $ in $ p' $ with  $ K_u $ given in \eqref{eq: state feedback gain}.  After $ p' $, the state trajectory $ s^* $ (an infinite path in $ \Go $) will keep repeating the MMC $ c^* $. Consequently, we don't care about the other columns of $ K_u $, because they correspond to states that will never be encountered. The correctness of the state-feedback gain $ K_u $ in \eqref{eq: state feedback gain} for the optimal control input sequence is thus proved. 
	
	We can prove the correctness of the other state-feedback gain matrix $ K_{\sigma} $ in \eqref{eq: state feedback gain} for the optimal switching signal in precisely the same way as that for $ K_u $ above. The proof of Theorem \ref{thm: state feedback} is therefore finished.
\end{IEEEproof}
\begin{remark}
	Eq. \eqref{eq: J u* w*} in proof of Theorem \ref{thm: optimal} implies that the cost of the transient path $ \pt $ in Theorem \ref{thm: state feedback} does not affect the optimal value $ J^* $. Thus, $ \pt $ can be any simple path from $ \dN^{i_0^*} $ to $ \dN^{i_{\alpha}^*} $ in $ \Go $, not necessarily the  shortest one (i.e., the one of the minimum weight), as long as it satisfies the condition in the first statement of Theorem \ref{thm: state feedback}. 
\end{remark}

\subsection{Efficient Algorithm Design} \label{sec: implementation}
Theorem \ref{thm: optimal} and Theorem \ref{thm: state feedback} have established the connection between the average-cost IHOC and the MMC in the associated OSTG. The remaining problem is how to locate an MMC in the OSTG such that the optimal state trajectory \eqref{eq: optimal trajectory} and the state-feedback gain matrices \eqref{eq: state feedback gain} can be constructed. One method is the exhaustive enumeration of all simple cycles \cite{zhao2010optimal}, which is only applicable to small networks. The start-of-the-art algorithm in terms of time efficiency to find such an optimal cycle in the input-state space is the Floyd-like algorithm first proposed in \cite{zhao2011floyd} and afterward applied to SBCNs in \cite{li2014optimal}, whose time complexity is still high though. As a major contribution of this study, we develop a more efficient method  by resorting to Karp's MMC algorithm \cite{karp1978characterization} in graph theory. Note that we focus on the OSTG of the SBCN rather than the much larger input-state space. 

Given a directed graph $ G =(V, E, w) $, let $ o \in V $ be a source vertex  that can reach every vertex in $ G $. Let $ F(k, v) $ be the minimum weight of any $ k $-edge path from $ o $ to  $ v \in V$. If no such path exists, $ F(k, v)\deq  \infty $. Karp proves that:
\begin{lemma} \label{lemma: Karp result} \cite{karp1978characterization}
	Supposing the minimum mean weight of cycles in \eqref{eq: MMC} is $ \mu^* = \bar{w}(c^*)$, it can be computed by
	\begin{equation} \label{eq: Karp}
		\mu^* = \min_{v \in V} \max_{0 \le k \le |V| - 1} \frac{F(|V|, v) - F(k, v)}{|V| - k}.
	\end{equation}
\end{lemma}
\begin{remark}
	If such a source vertex $ o $ does not exist in $ G $, then we can first partition $ G $ into several strongly connected components (SCCs), and then find the MMC in each SCC \cite{karp1978characterization}. Nonetheless, this is not a problem in our case, since we can always choose $ x_0 $ as $ o $ for our OSTG. 
\end{remark}

Lemma \ref{lemma: Karp result} identifies the minimum mean weight $ \mu^* $, but it does not state how to pinpoint such an MMC.  Though Karp mentioned the construction of an MMC roughly in \cite{karp1978characterization}, a very recent paper \cite{CHATURVEDI201721} spots an error in his procedures and gives a correct one instead as follows.

\begin{lemma} \label{lemma: MMC finding} \cite{CHATURVEDI201721}
	Let $ v^* $ and $ k^* $ be an optimal solution to \eqref{eq: Karp}. Every cycle on the $ |V| $-edge path from $ o $ to $ v^* $ of weight $ F(|V|, v^*) $ is an MMC, where $ o $ is the source vertex.
\end{lemma}

Let's come back to the OSTG. Combining the above two lemmas with Theorem \ref{thm: state feedback}, we have the following conclusion.
\begin{theorem} \label{thm: trajectory construction}
	Consider Problem \ref{prob: 1} and its OSTG $ \Go = (V, E, w, x_0) $ with $ x_0 = \dN^{i_0} $. Choose $\dN^{i_0} $ as the source vertex $ o $, and let $ v^*$ and $ k^* $ be an optimal solution to \eqref{eq: Karp} for $\Go $. \rv{Suppose a $ |V| $-edge path from $ o $ to $ v^* $ in $ \Go $ is $ p^* = \dN \big< i_0^*, i_1^*, \cdots, i_{|V|}^* \big>$, where $ \dN^{i_{|V|}^*} = v^* $ and $ i_0^* = i_0 $ by construction. Let the first simple cycle in $ p^* $ be $ c^* = \dN\big< i_{\alpha}^*, i_{\alpha+1}^*, \cdots, i_{\beta}^*,   i_{\alpha}^* \big> $, which is preceded by a sub-path $ \pt = \dN\big< i_0^*, i_1^*, \cdots, i_{\alpha}^*\big>$. With $ c^* $ and $ \pt $, the optimal state trajectory $ s^* $ in Theorem \ref{thm: state feedback} can be constructed by \eqref{eq: optimal trajectory}.}
\end{theorem}
\begin{IEEEproof}
	First, the existence of $ p^*$ can be proved easily by contradiction: if $ p^* $ does not exist, then $ F(|V|, v^*)  = \infty$, which indicates $ v^* $ cannot be an optimal solution to \eqref{eq: Karp}. 
	
	Next, since $ p^* $ has $ |V|+1 $ vertices, there must exist cycles in $ p^* $, which further implies the existence of simple cycles because a non-simple cycle can be decomposed into simple ones by Proposition \ref{prop: decomposition}. The first simple cycle $ c^* $ can be easily identified by a linear scan of $ p^* $ (see Line \ref{Line: start c^*} -- \ref{Line: end c^*} in Algorithm \ref{alg: main}). Lemma \ref{lemma: MMC finding} ensures that $ c^* $ must be an MMC. Therefore, $ c^* $ is a simple MMC. Since $ c^* $ is the first simple cycle, there is no overlap between the transient sub-path $ \pt $ and  the MMC $ c^* $, which conforms to all  requirements of $ s^* $ in \eqref{eq: optimal trajectory}.
\end{IEEEproof}

We have finished all the theoretical work at this point. Theorem \ref{thm: trajectory construction} indicates an algorithm composed of three tasks:
(i) solve \eqref{eq: Karp} to get $ v^* $; (ii) find a $ |V| $-edge path $ p^* $ from $ o $ (i.e., $ x_0 $) to $ v^* $ of weight $ F(|V|, v^*) $; (iii) build the state trajectory $ s^*$ \eqref{eq: optimal trajectory} using $ p^* $. Given the OSTG $ \Go = (V, E, w, x_0) $, we can solve \eqref{eq: Karp} in Task (i) highly efficiently via dynamic programming (DP) \cite{karp1978characterization, cormen2009introduction} based on the recursion below:
\begin{equation}
	F(k+1, \dN^j) = \min_{(\dN^i, \dN^j) \in E}  F(k, \dN^i) + w(\dN^i, \dN^j), \  k \ge 0
\end{equation}
and the base case:
\begin{equation} \label{eq: base case}
	F(0, \dN^i) = \begin{cases}
	0, \  \dN^i = x_0 \\
	\infty, \ \textrm{otherwise}
	\end{cases}.
\end{equation}
Task (ii) can be finished simultaneously with Task (i) by keeping track of the vertices in a path with a backpointer. After that, Task (iii) is straightforward to be completed, and it is trivial to get the state-feedback gains \eqref{eq: state feedback gain} from $ s^* $. We detail the procedures for the three tasks in Algorithm \ref{alg: main}.

\begin{algorithm}[htb]
	\caption{State-Feedback Control Design using the OSTG} \label{alg: main}
	\begin{algorithmic}[1] 
		\Input The OSTG $ \Go = (V, E, w, x_0) $ of Problem \ref{prob: 1}, $ x_0 = \dN^{i_0} $
		\Output The state-feedback gain matrix $ K_u $ and $ K_{\sigma} $ in \eqref{eq: state feedback law}
		\State Initialize $ (N+1) \times (N+1) $  arrays $ F $ and $ B $ with $ \infty $  \label{Line: arrayBD}
		\State $ F[0, i_0] \gets 0$ (see \eqref{eq: base case})
		\LineComment Task (i)
		\ForAll{$ k \gets 1 $ to $ |V| $}  \label{line: for Karp outside begin}
		\ForAll{$ \dN^j \in V $}  \label{line: for Karp begin}
		\State $ F[k, j] \gets \min_{(\dN^i, \dN^j) \in E}  F[k - 1, i] + w(\dN^i, \dN^j) $ \label{line: dp min}
		\State $ B[k, j] \gets i^*$, where $ i^* $ is the minimizer in Line \ref{line: dp min}
		\EndFor \label{line: for Karp end}
		\EndFor  \label{line: for Karp outside end}
		\State Solve \eqref{eq: Karp} by enumerating $ F $ and get $ v^* \gets \dN^{i^*_{|V|}} $ \label{line: find min}
		\LineComment Task (ii)
		\State Create an array $ p^* $ of size $ |V| + 1 $ with $ p^*[|V|] \gets  i^*_{|V|} $
		\ForAll{$ k \gets |V|$ to $ 1 $}
		\State $ p^*[k - 1] \gets B[k, p^*[k]] $
		\EndFor
		
		\LineComment Task (iii)
		\State Initialize an integer array $ A $ of size $ N+1 $ with -1  \label{Line: start c^*} \label{Line: arrayA}
		\ForAll{$ t \gets 0 $ to $ |V| $}  \label{line: find MMC begin}
		\State $ i_t^* \gets p^*[t] $
		\If{$ A[i_t^*] = -1$}  
		\State $ A[i_t^*] \gets t $
		\Else \Comment The first simple cycle is found
		\State $ \alpha \gets A[i_t^*], \quad \beta \gets t - 1 $   \Comment See Theorem \ref{thm: trajectory construction}
		\State \textbf{break}
		\EndIf
		\EndFor  \label{Line: end c^*}
		\State The state trajectory in \eqref{eq: optimal trajectory} is   \label{Line: s^*}
		\begin{align*}
		s^* \gets \dN(&p^*[0], p^*[1], \cdots, p^*[\alpha], p^*[\alpha+1], \cdots, p^*[\beta], \nonumber\\
		&p^*[\alpha], p^*[\alpha+1], \cdots, p^*[\beta], \cdots)
		\end{align*}
		\State Compute $ K_u $ and $ K_{\sigma} $ by \eqref{eq: state feedback gain} with $ s^* $  \label{line: 27}
	\end{algorithmic}
\end{algorithm}

\begin{remark}  \label{rmk: save memory with hash}
	The arrays in Line \ref{Line: arrayBD} and Line \ref{Line: arrayA} of Algorithm \ref{alg: main} can be replaced by dictionaries (i.e., hash tables \cite{cormen2009introduction}), without affecting time complexity, to save memory space in practice, because we have $ |V| < N$ or even $ |V| \ll N $ in most cases.
\end{remark}

\textit{Time Complexity Analysis:} In Algorithm \ref{alg: main}, the Task (i) part solves \eqref{eq: Karp}  via DP by computing each $ F(k, v) $  only once. Given a fixed $ k $, the inner loop (Line \ref{line: for Karp begin} -- \ref{line: for Karp end}) visits all $ |E| $ edges in $ \Go $ once. Thus, the full loop (Line \ref{line: for Karp outside begin} -- \ref{line: for Karp outside end}) runs in time $ O(|V||E|) $. Line \ref{line: find min} takes $ O(|V|^2) $ further operations to find the minimizer. Next, Task (ii) requires $ O(|V|) $ operations. Finally, in Task (iii), it is obvious that  the loop (Line \ref{line: find MMC begin} -- \ref{Line: end c^*}), Line \ref{Line: s^*}, and Line \ref{line: 27} all run in $ O(|V|) $. 

Since each vertex in $ \Go $ has at most $ zM $ outgoing edges (see Lemma \ref{lemma: successor}), we have $ O(|E|) = O(zM|V|) $. Thus, the overall running time is dominated by Task (i), which is $ O(|V||E|) = O(zM|V|^2) $. Furthermore, due to $ |V| \le N $, the worst-case time complexity of Algorithm \ref{alg: main} is $ O(zMN^2) $. Recall that Algorithm \ref{alg: OSTG} takes $ O(zMN) $ time to build the OSTG $\Go $. Combing the two algorithms, the overall time complexity of our graph-theoretical approach is $ O(zMN^2) $.

\begin{example} \label{ex: solve toy problem}
	Recall Example \ref{example: 1} and its OSTG $ \Go $ in Fig. \ref{fig: OSTG of example 1}.  Algorithm \ref{alg: main} generates the following results for Theorem \ref{thm: trajectory construction}:
	\begin{itemize}
		\item One vertex in $ \Go $ that minimizes \eqref{eq: Karp} is $ v^* = \delta_{8}^2 $ along with $ k^* = 3 $. We have $ \mu^* $ = 3.5  and $ F(7, \delta_{8}^2) = 29$.
		\item A $ 7 $-edge path from $ o = \delta_{8}^1 $ to $ v^* $ of weight $ F(7, \delta_{8}^2) $ is $ p^* = \delta_{8}\left< 1, 7, 8, 2, 5, 7, 8, 2 \right>. $
		\item The first simple cycle in $ p^* $ is delimited by $ \alpha = 1 $ and $ \beta = 4 $, which yields an MMC $ c^* = \delta_{8} \left< 7, 8, 2, 5, 7 \right> $, whose mean weight is exactly $ \bar{w}(c^*) = 3.5 $. In other words, the SBCN \eqref{eq: example 1 model} will converge to an \textit{attractor} \cite{zhao2010input} $ c^* $ under IHOC after just 1 step.
		\item The optimal infinite state trajectory \eqref{eq: optimal trajectory} is therefore $ s^* = \delta_{8}(1, 7, 8, 2, 5, 7, 8, \cdots)$ and the feedback gains \eqref{eq: state feedback gain} are $ K_u = \delta_2 [1\; 2 \; * \; * \; 2\; *\; 2\; 2], \ K_{\sigma} = \delta_2 [1\; 1 \; * \; *\; 1\; *\; 1\; 2], $ 
		where $ * $ indicates that this column can be arbitrarily set.
	\end{itemize}
\end{example}

\section{Comparison with Existing Methods} \label{sec: time complexity}
A primary challenge in IHOC of large-scale BCNs (SBCNs) is the prohibitively high computational cost \cite{wu2019optimal}. In this section, we compare the proposed graph-theoretical approach with existing methods in respect of time complexity to highlight its superior efficiency. Besides, we introduce a simple technique that optimizes some existing methods to reduce their running time given a specific initial state in practice for fair comparison later in the next section  (i.e., Section \ref{sec: benchmark}).

\subsection{Conversion between BCNs and SBCNs} \label{sec: conversion}
As reviewed in Section \ref{sec: intro}, most work on IHOC with average cost focuses on non-switched BCNs. To the best of our knowledge, only Ref. \cite{li2014optimal} considers SBCNs. Nonetheless, since a normal BCN is just a special SBCN with a single subsystem, i.e., $ z=1 $, both our method and the one in \cite{li2014optimal} can be applied directly to BCNs. More interestingly, the opposite is also true: an SBCN can also be transformed into a normal BCN. In \cite{chen2014output}, an SBCN with a stationary state-dependent switching law is converted to a non-switching BCN. We derive a similar result for time-dependent SBCNs in this study via control input augmentation as follows. 

Set $ \bar{L} \deq [L_1, L_2, \cdots, L_z] \in \cL_{N\times zMN}$ as an augmented network transition matrix. The SBCN \eqref{eq: ASSR-SBCN} is thus equivalently expressed by a BCN as,
\begin{align}\label{eq: BCN equivalence}
x(t+1) =  \bar{L}\sigma(t)u(t)x(t) = \bar{L}\bar{u}(t)x(t),		  
\end{align}
where $ \bar{u}(t) \deq \sigma(t)u(t) \in \Delta_{zM}$ is an augmented control input that integrates both control and switching signals.

Algorithms initially developed for optimal control of BCNs can thus be applied to SBCNs in the form \eqref{eq: BCN equivalence}, though we target SBCNs directly. The transformation between BCNs and SBCNs makes it reasonable to compare the performance of our approach with existing methods originally devised for BCNs.

\subsection{Time Complexity Comparison} \label{sec: time complexity comparison}
Since most existing methods are developed for BCNs, we assess the time complexity in this section by applying our approach to a BCN, i.e., a single-subsystem SBCN. 
In the literature, the average-cost IHOC problem of BCNs was first investigated in \cite{zhao2010optimal} and later considered in \cite{zhao2011floyd,fornasini2013optimal,wu2019optimal}. By convention, the worst-case time complexity is used to indicate the efficiency of algorithms \cite{fornasini2013optimal,cormen2009introduction,karp1978characterization}. We list the (worst-case) time complexity of existing methods for BCNs in a chronological order in Table \ref{tbl: time complexity}. Recall that $ N \deq 2^n $ and $ M \deq 2^m $, where $ n $ and $ m $ refer to the number of nodes and  control inputs in a BCN respectively.

\begin{table*}[tb!]
	\centering
	\caption{Time Complexity Comparison for IHOC of BCNs with Average Cost}
	\label{tbl: time complexity}
	\begin{tabular}{lcccccc} 
		\toprule
		Method          & Cycle enumeration \cite{zhao2010optimal}     &Floyd-like \cite{zhao2011floyd}   &Value iteration \cite{fornasini2013optimal}   &Floyd-like \cite{li2014optimal}   &Policy iteration \cite{wu2019optimal}   & Proposed   \\ 
		\hline
		Time complexity & $O(M^{2N-1}N^3)$ & O($M^4N^4$) & $- $ (unbounded) & $ O(MN + N^4)$ & $O(M^N (N^2+MN))$ & $O(MN^2)$  \\
		\bottomrule
	\end{tabular}
\end{table*}

The first method \cite{zhao2010optimal} in Table \ref{tbl: time complexity} evaluates all cycles of length ranging from $ 1 $ to $ MN $ in the input-state space to locate the MMC. This brute-force method will quickly become intractable as $ n $ and $ m $ increase. An immediate improvement is the Floyd-like algorithm \cite{zhao2011floyd} adapted from the Floyd-Warshall algorithm \cite{cormen2009introduction} in graph theory, which essentially still enumerates all cycles like \cite{zhao2010optimal}  but more economically via dynamic programming. The Floyd-like algorithm is later applied to SBCNs in \cite{li2014optimal}, but the cycles are found in the state space instead of the larger input-state space, which helps reduce the computational complexity. The other two work approaches this problem from a different angle by putting it in a value iteration \cite{fornasini2013optimal} or policy iteration \cite{wu2019optimal} framework. However, as pointed out by \cite{wu2019optimal}, the value iteration approach in \cite{fornasini2013optimal} is not guaranteed to converge to the optimal solution in finite steps. Though the policy iteration approach \cite{wu2019optimal} can yield a stationary state-feedback optimal policy in finite iterations, its worst-case time complexity is extremely high.  

When interpreting the time complexity in Table \ref{tbl: time complexity}, note that we assume $ M \le N $ for BCNs (or $ zM \le N $ for SBCNs), because a state can transit to at most $ N $ succeeding states regardless of the number of inputs. \rv{That is, one vertex in the OSTG has no more than $ N $ outgoing edges irrespective of how large $ M$ is. Moreover, one interesting observation is that we can control the whole network by manipulating only a fraction of the nodes \cite{zhong2018pinning, murrugarra2016identification}. Consequently, we typically have $ m < n $ and thus $ M \ll N $ in practice especially for large networks.} In Table \ref{tbl: time complexity}, a more concise measure of the time complexity of our approach is $ O(N^3) $, while the previously state-of-the-art Floyd-like algorithm \cite{li2014optimal} runs in $ O(N^4) $. The reduction of running time that our approach achieves can be striking when handling large networks, because $ N = 2^n $ can be considerably large. \rv{As a side note, if we do not make any assumption about the relative size of $ M $ and $ N $, the precise time complexity of our approach for a BCN is $ O(MN + \min(MN^2, N^3)) $ instead, since there are at most $ \min(MN, N^2) $ edges in the OSTG even if $ M > N $. However, as aforementioned, it is reasonable to assume $ M \le N $ from a practical standpoint, which is adopted throughout this paper to ease discussion without affecting the conclusion: our approach has the lowest time complexity. }

As discussed in Section \ref{sec: conversion}, the methods listed in Table \ref{tbl: time complexity} are also applicable to SBCNs in form \eqref{eq: BCN equivalence}, and their time complexity can be obtained simply by replacing $ M $ with $ zM $. Simple calculations will lead to the same conclusion: our proposed approach achieves the highest time efficiency even if we view an SBCN as a BCN with augmented control inputs. Finally, note that, though the Floyd-like algorithms in \cite{zhao2011floyd} and \cite{li2014optimal} borrow ideas from graph theory as well, they still operate on large matrices rather than dedicated graphs like our OSTG.

\begin{remark}
	Despite its prohibitively high worst-case time complexity, the policy iteration approach may generally converge to the optimal solution in a few iterations  \cite{wu2019optimal}. Even so, our approach still ran much faster when tested with the Ara operon network considered in \cite{wu2019optimal} (see Section \ref{sec: benchmark}).
\end{remark}

\begin{remark} \label{rmk: exponential}
\rv{T. Akutsu \textit{et al. } have proved that control problems for general BCNs are NP-hard \cite{akutsu2007control}. That is, a polynomial-time algorithm does not exist for such problems unless $ P=NP $, while $ P \ne NP $ is a widely believed conjecture. More details on NP-completeness can be found in \cite[Chapter 34]{cormen2009introduction}. This fact rationalizes the exponential-time algorithms on BCNs in the literature, whose time complexity is generally in a polynomial of $ N = 2^n $. Consequently, the aim of our study is not to develop polynomial-time algorithms, which is still a fundamental unsolved problem in computer science today. Nevertheless, the intimidating NP-hardness does not necessarily eliminate the possibility of further enhancing algorithm efficiency, for example, to decrease the degree of the polynomial in $ N $, which is exactly our attempt in this paper. Finally, we acknowledge that, though the proposed approach can handle relatively larger networks that are beyond the capacity of existing methods, optimal control of huge networks, like those with hundreds of nodes, is still intractable and remains an open problem. To deal with such large networks in practice, we may resort to approximation algorithms like \cite{fornasini2013optimal} to get an approximate solution or make use of the special structure of a network, if applicable, to reduce a large network into smaller ones and then apply an \textit{divide and conquer} strategy\cite{zhao2015control}.}
\end{remark}

\subsection{Extension to All Initial States} \label{sec: extension}

\rv{
One may notice that the proposed approach actually solves the IHOC problem for one specific initial state, i.e., $ x_0 $ in Problem \ref{prob: 1}, to get the optimum $ J^*(x_0) $. This judgment is also true for prior work \cite{zhao2010optimal} and \cite{zhao2011floyd} that essentially deals with the reachable set $ \cR(x_0) $ for a BCN. The later study \cite{li2014optimal} presents a straightforward adaptation of \cite{zhao2011floyd} to handle SBCNs but only detects the MMC (called \textit{optimal cycle} therein) in the complete state space $ \DN $. That is, the algorithm in \cite{li2014optimal} obtains directly the optimal objective value among all initial states, $ J^{**} = \min_{x \in \DN} J^*(x) $, but it cannot tell $ J^*(x_0) $ for a specific initial state $ x_0 $. Note that our algorithm can also be applied to an OSTG with $ V=\DN $ to easily get $ J^{**} $. By contrast, both the value iteration \cite{fornasini2013optimal} and policy iteration \cite{wu2019optimal} based methods solve the problem for all initial states of a BCN simultaneously. However, the former only yields an approximate solution, while the worst-case time complexity of the latter in Table \ref{tbl: time complexity} is extremely high (apart from its high memory consumption shown in Section \ref{sec: benchmark}). Now the question is how we can extend the proposed approach to solve the optimal control problem for each initial state  instead of a particular $ x_0 $ while preserving its efficiency.}

\rv{
A naive fix is to run the proposed algorithm $ N $ times, each from one of the $ N $ initial states (or $ C_x $ only if constraints are applied). For a BCN, the naive routine leads to time complexity $ O(MN^3) $ that is still at least as good as the state-of-the-art one $ O(MN + N^4) $ \cite{li2014optimal} (Table \ref{tbl: time complexity}), though only one (not explicitly specified) initial state is considered in \cite{li2014optimal}. Nevertheless, we notice that, in general, more than one state (vertex) can reach the same MMC, and we should avoid repetitive computation. An optimized \textit{recursive elimination} procedure in a backward manner is designed as follows: first detect the MMC, and then get all states that reach this MMC. 
\begin{enumerate}
	\item Build a complete OSTG $ G $ via Algorithm \ref{alg: OSTG} but starting from every \textit{unvisited} initial state in $ C_x $.
	\item Find an MMC $ c^* $ of $ G $ by Lemma \ref{lemma: Karp result} and Lemma \ref{lemma: MMC finding}. Collect all states (vertices) in $ G $ that can reach $ c^* $ into a set $ S $. Problem \ref{prob: 1} is thus easily solved for each initial state $ x_0 \in S $ by trivially adapting Algorithm \ref{alg: main}.
	\item Eliminate the vertices in $ S $ and related edges from $ G $ to get a sub-graph $ G' $. Let $ G \leftarrow G' $.
	\item Go to step 2) until $ G $ is empty.
\end{enumerate}
}

\rv{
\textit{Time Complexity Analysis:} Step 1) still runs in linear time $ O(zMN) $ via a BFS principle (see Algorithm \ref{alg: OSTG}), and the set $ S $ in Step 2) can be obtained easily by reversing the edge directions of $ G $ in time $ O(zMN) $ as well. Besides, finding the MMC of a complete OSTG in the first iteration still runs in $ O(zMN^2) $, since there are at most $ N $ vertices and $ zMN $ edges. Step 3) runs obviously in linear time as well. The rationality of Step 3) is that, any remaining vertex (state) $ x' \in G' $ cannot have a state trajectory  that passes through any vertex $ x \in  S $; otherwise, $ x' $ could reach the MMC $ c^* $ through $ x $ in Step 2), which forms a contradiction. As a result, the elimination of $ S $ from $ G $ will not affect the optimal state trajectory for any initial state $ x' \in G' $. Though it is hard to compute the precise time complexity of the above procedure, it is definitely much lower than the complexity of the naive method, since the size of the graph keeps decreasing and the number of iterations between Step 2) and 4), termed $ K $, is generally much smaller than $ N $. An excessively loose upper bound of the time complexity is thus $ O(zKMN^2) $ with $ K \le N $ and typically $ K \ll N $. In summary, if needed, our method can be extended to solve the optimal control problem for each initial state even with lower time complexity than the state-of-the-art algorithm \cite{li2014optimal} that only deals with one particular initial state. It is also evident that the above method has much lower time complexity than the only existing work \cite{wu2019optimal}  that truly handles all initial states (see Table \ref{tbl: time complexity}).
}

\section{A Benchmark Example: Optimal Intervention in T-LGL Leukemia} \label{sec: benchmark}
\begin{figure}[tb!]
	\centering
	\includegraphics[width=60mm]{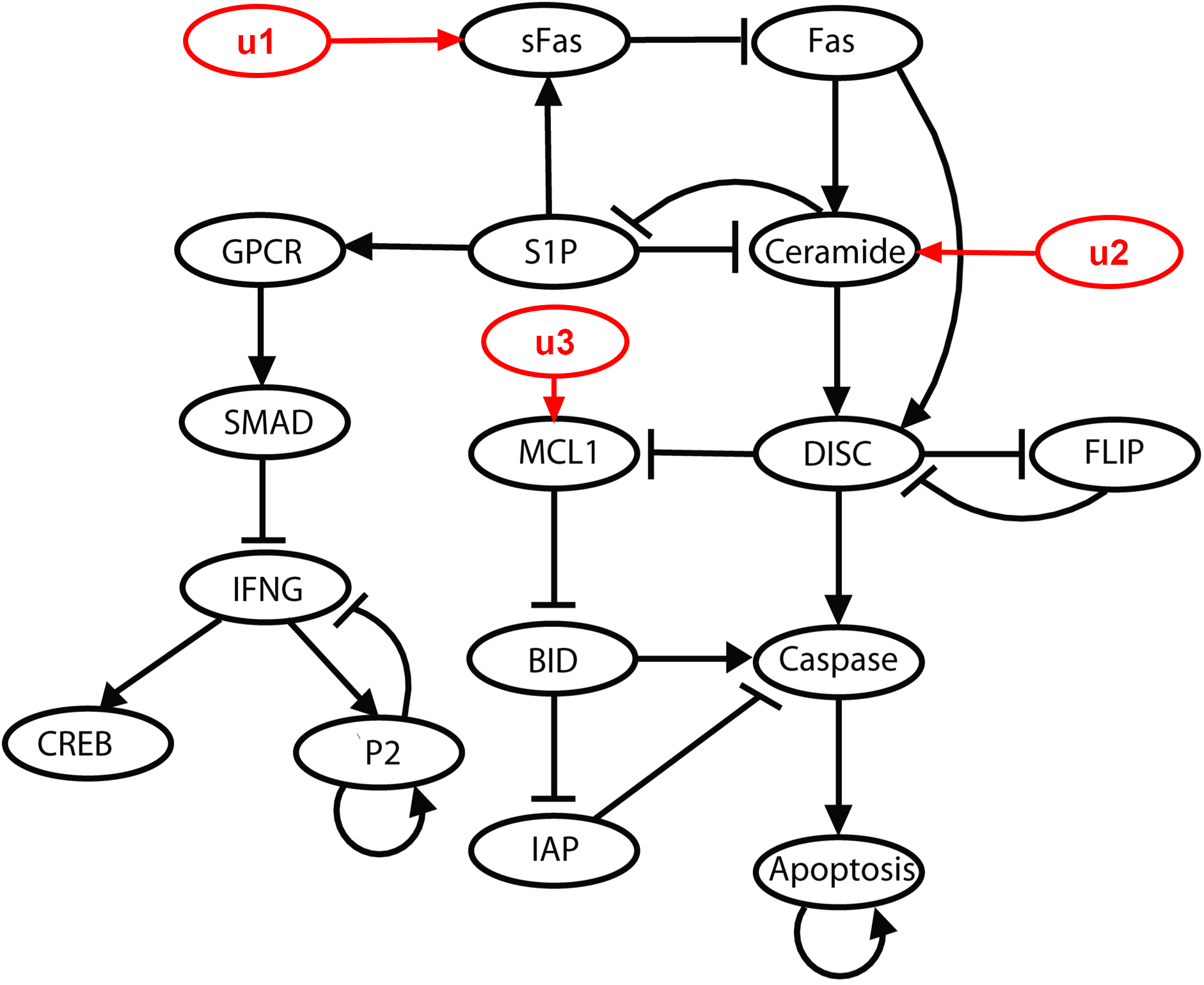} 
	\caption{The reduced T-LGL signaling network with external control (adapted from \cite{saadatpour2011dynamical}). Sharp and hammerhead arrows denote activation and inhibition respectively. The inhibitory edges from Apoptosis to other nodes are
		not shown for clarity. The red circles and arrows indicate the external control.}
	\label{fig: T-LGL}
\end{figure}

Most existing theoretical studies on BCNs only deal with tiny networks, typically comprising no more than 5 nodes, for illustration purpose.  To benchmark our approach against existing ones, we use a signaling network  in blood cancer, the \textit{T cell large granular lymphocyte} (\textit{T-LGL}) \textit{leukemia}, a chronic disease characterized by an abnormal increase of cytotoxic T cells \cite{saadatpour2011dynamical}.  This network includes 16 nodes, as shown in Fig. \ref{fig: T-LGL}.  One possible treatment of such diseases is to apply external intervention to force the activation or inhibition of specific nodes in a network through drugs, radiation, or chemo \cite{murrugarra2016identification}. In this section, we aim to steer the T-LGL network from a diseased state and maintain it at a healthy state using IHOC like \cite{pal2006optimal}. \rv{Interested readers may refer to \cite{saadatpour2011dynamical} and \cite{murrugarra2016identification} for more details about the biological background of the T-LGL network.}

Section \ref{sec: conversion} has shown that a BCN and an SBCN are interchangeable via mathematical manipulations. The non-switching T-LGL network is purposefully chosen here  to suit existing methods originally developed for BCNs, like \cite{zhao2011floyd}, \cite{fornasini2013optimal} and \cite{wu2019optimal}. Its Boolean functions are listed in Table \ref{tbl: T-LGL} \cite[Table S3]{saadatpour2011dynamical}. This network has also been studied in \cite{murrugarra2016identification} to identify potential intervention targets.  Supposing we apply intervention to sFas, Ceramide, and MCL1 (indicated by red arrows in Fig. \ref{fig: T-LGL}), we get a BCN including 16 state variables and 3 control inputs, i.e., $ N = 65536 $ and $ M = 8 $. Following the problem setting in \cite{murrugarra2016identification}, the network is initially in a diseased state $0001101000101110$ (i.e., $x_0 = \delta_{65536}^{58834}$), where Caspase and Apoptosis are OFF, and we want to drive it to a healthy state $ 0000000000000001 $ (i.e., $x_h = \delta_{65536}^{65535}$) with Apoptosis activated. No constraints are set here because most existing methods do not handle constraints in their algorithms.

\begin{table}[tb]
	\centering
	\caption{Boolean Rules of the T-LGL Boolean Control Network}
	\label{tbl: T-LGL}
	\begin{tabular}{ll} 
		\toprule
		Node      & Boolean rule                                        \\ 
		\midrule
		CREB      & IFNG  $\land$  $\lnot$Apoptosis                              \\
		IFNG      & $\lnot$(SMAD $\lor$ P2 $\lor$ Apoptosis)                       \\
		P2        & (IFNG $\lor$ P2) $\land$ $\lnot$Apoptosis                      \\
		GPCR      & S1P $\land$ $\lnot$Apoptosis                               \\
		SMAD      & GPCR $\land$ $\lnot$Apoptosis                              \\
		Fas       & $\lnot$(sFas $\lor$ Apoptosis)                             \\
		sFas      & S1P $\land$ $\lnot$Apoptosis $\land$ $ u_1 $                        \\
		Ceramide  & Fas $\land$ $\lnot$(S1P $\lor$ Apoptosis) $\lor$ $ u_2 $                \\
		DISC      & (Ceramide $\lor$ (Fas $\land$ $\lnot$FLIP)) $\land$ $\lnot$Apoptosis  \\
		Caspase   & ((BID $\land$ $\lnot$IAP) $\lor$ DISC) $\land$ $\lnot$Apoptosis       \\
		FLIP      & $\lnot$(DISC $\lor$ Apoptosis)                             \\
		BID       & $\lnot$(MCL1 $\lor$ Apoptosis)                             \\
		IAP       & $\lnot$(BID $\lor$ Apoptosis)                              \\
		MCL1      & $\lnot$(DISC $\lor$ Apoptosis) $\land$ $ u_3 $                      \\
		S1P       & $\lnot$(Ceramide $\lor$ Apoptosis)                         \\
		Apoptosis & Caspase $\lor$ Apoptosis                                \\
		\bottomrule
	\end{tabular}
\end{table}

To achieve the above objective, we set up a simple stage cost function for Problem \ref{prob: 1} as follows:
\begin{equation} \label{eq: cost T-LGL}
	g(x(t), u(t), \sigma(t)) = \begin{cases}
	1, \ x(t) = x_h \\
	5, \ \textrm{otherwise}
	\end{cases}.
\end{equation}
Note that a biologically reasonable cost function must be designed by domain experts in practice \cite{pal2006optimal}. We adopt \eqref{eq: cost T-LGL} mainly for quick verification of the algorithms' correctness: obviously,  an optimal strategy should finally pin the network to the fixed point $ x_h $ with the optimal value  $ J^* = 1$.  

\begin{figure}[tb]
	\centering
	\includegraphics[width=70mm]{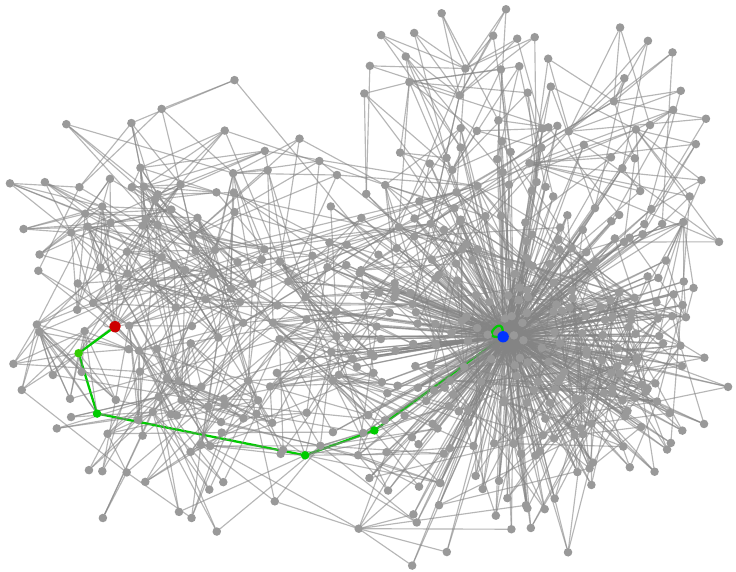} 
	\caption{The OSTG of the T-LGL network and the optimal state trajectory yielded by our approach. The red circle and the blue circle denote the initial state $ x_0 $ and the desired state $ x_h $ respectively. The state trajectory steered by the optimal policy is indicated by green lines. Note the MMC around $ x_h $.}
	\label{fig: T-LGL-OSTG}
\end{figure}

Applying our Algorithm \ref{alg: OSTG} and Algorithm \ref{alg: main} in turn, we get the following results easily in only about 3.5 seconds.
\begin{itemize}
	\item The OSTG starting from $ x_0 $ has only $ |\cR(x_0)| = 468 $ vertices in total (Fig. \ref{fig: T-LGL-OSTG}), though the full state space has up to $N = 65536 $ states. This fact justifies our previous analysis:  there usually exists $ |\cR(x_0)|  \ll N $ for a large-scale network with a small number of control inputs.
	\item The minimum cycle mean is $ \mu^* = 1 $, which is obtained by $ v^* = \delta_{65536}^{65279} $ and $ k^* = 5 $ in \eqref{eq: Karp}. 
	\item A 468-edge path from $ x_0 $ to $ v^* $ in Theorem \ref{thm: trajectory construction} is
	\begin{align*}
		p^* = \delta_{65536} \big< &58834, 59094, 58184, 62126, 60175,  \\
			&\underbrace{65535, \cdots, 65535}_{463}, 65279   \big>.
	\end{align*}
	\item  The first simple cycle in $ p^* $ is delimited by $ \alpha = 5$ and $ \beta =5  $ in Theorem \ref{thm: trajectory construction}, i.e., $ c^* =  \delta_{65536} \big< 65535, 65535 \big>$, whose mean weight is exactly 1. Recall that the desired destination state is $ x_h = \delta_{65536}^{65535} $. The optimal trajectory 
	\begin{align} \label{eq: optimal trajectory of T-LGL}
		s^* = \delta_{65536} ( &58834, 59094, 58184, 62126, 60175,  \nonumber\\
			&65535, 65535, 65535,\cdots ) 
	\end{align}
	 converges to the desired fixed point $ x_h $, exactly as we have expected, after 5 steps driven by the optimal policy, which is illustrated in Fig. \ref{fig: T-LGL-OSTG}.
	 \item The optimal state-feedback control law in \eqref{eq: state feedback gain} is
	 \begin{align*}
	 	\textrm{Col}_{58834}(K_u) = \delta_{8}^{6}, \ \textrm{Col}_{59094}(K_u) = \delta_{8}^{8}, \\
	 	\textrm{Col}_{58184}(K_u) = \delta_{8}^{1}, \ \textrm{Col}_{62126}(K_u) = \delta_{8}^{8}, \\
	 	\textrm{Col}_{60175}(K_u) = \delta_{8}^{3}, \ \textrm{Col}_{65535}(K_u) = \delta_{8}^{3},
	 \end{align*}
	 and no switching is required for this normal BCN.
	 \item Multiple runs show that Algorithm \ref{alg: OSTG} takes about 2.2 s and Algorithm \ref{alg: main} takes around 1.3 s. 
\end{itemize}

To verify the efficiency of our graph-theoretical approach, we also try other methods in Table \ref{tbl: time complexity} to solve this optimal control problem and measure their running time. The brute-force cycle enumeration method \cite{zhao2010optimal} is skipped because of its obvious incapacity for this relatively large-scale problem.  Since the Floyd-like algorithm in  \cite{li2014optimal} improves the original one proposed in \cite{zhao2011floyd} by operating in the state space instead of the input-state space to reduce computational complexity (Table \ref{tbl: time complexity}), we use the former as a representative in this benchmark test. As for the value iteration approach \cite{fornasini2013optimal} , the number of iterations needed depends on the accuracy we want, and it may never get the exact optimal solution in finite steps, as revealed by \cite{wu2019optimal}. We thus measure its running time to obtain $ \epsilon$-suboptimal solutions. \rv{Note that we have evaluated the Floyd-like algorithm \cite{li2014optimal} and the value iteration approach \cite{fornasini2013optimal} in both their original version and optimized version for fair comparison. We consider only $ \cR(x_0) $ in the latter version instead of the complete state space $ \DN $ in the former, though both methods \cite{li2014optimal, fornasini2013optimal} operate originally on $ \DN $ that introduces unnecessary complexity.} Unfortunately, such optimization cannot be easily applied, if possible, to the policy iteration approach \cite{wu2019optimal} due to its increased sophistication.

\begin{table*}
	\centering
	\caption{Running Time  of Various Methods to Solve the IHOC Problem for the T-LGL Network}
	\label{tbl: benchmark time}
	\begin{threeparttable}
		\begin{tabular}{clcccccc} 
			\toprule
			\multicolumn{2}{c}{\multirow{2}{*}{Method~}} & \multirow{2}{*}{Floyd-like \cite{li2014optimal}} & \multicolumn{3}{c}{Value iteration \cite{fornasini2013optimal}}                                     & \multirow{2}{*}{Policy iteration \cite{wu2019optimal}}                                       & \multirow{2}{*}{Proposed}  \\ 
			\cmidrule{4-6}
			\multicolumn{2}{c}{}                         &                             & $\epsilon=0.1$ & $\epsilon=0.01$ & \multicolumn{1}{l}{$\epsilon=0.001$} &                                                                         &                            \\ 
			\midrule
			\multirow{2}{*}{Time} & Optimized           & 18680 s                     & 420 s          & 4213 s          & 42200 s\tnote{*}                              & \multirow{2}{*}{\begin{tabular}[c]{@{}c@{}}Out of\\memory\end{tabular}} & \multirow{2}{*}{3.5 s}     \\ 
			\cmidrule{2-6}
			& Original            & 132000 h\tnote{*}                    & 54460 s\tnote{*}        & 544600 s\tnote{*}        & 5446000 s\tnote{*}                            &                                                                         &                            \\
			\bottomrule
		\end{tabular}
		\begin{tablenotes}
			\item[*] Estimated running time, which is obtained by multiplying the number of iterations and the time per iteration.
		\end{tablenotes}
	\end{threeparttable}
\end{table*}

A desktop PC with a 3.4 GHz Core i7-3770 CPU, 16 GB RAM, and 64-bit Windows 10 is used. All algorithms are implemented using Python 3.7. The running time of each method is listed in Table \ref{tbl: benchmark time}, except the policy iteration approach \cite{wu2019optimal}, which runs out of memory due to its manipulation of huge matrices. We first notice from Table \ref{tbl: benchmark time} that the impact of the straightforward optimization can be indeed significant. For example, each iteration of the Floyd-like algorithm \cite{li2014optimal,zhao2011floyd} operating on the whole state space takes more than 2 hours, and it needs totally 65535 iterations; by contrast, it takes about 40 s per iteration after optimization with only 467 iterations in total.  
Table \ref{tbl: benchmark time} also highlights that our proposed approach takes a remarkably shorter time to acquire the exact optimal solution, thousands of times faster than the Floyd-like algorithm and the value iteration approach even after their optimization. 

Both our method and the Floyd-like algorithm can acquire the exact optimal value $ J^*=1 $, while the value iteration approach can only approximate $ J^* $ as the number of iterations increases. Noticing the 5-step optimal state trajectory \eqref{eq: optimal trajectory of T-LGL} and the simple stage cost function \eqref{eq: cost T-LGL}, we can derive the approximated optimal value yielded by the value iteration approach \cite{fornasini2013optimal} after $ T $ iterations as $ \tilde{J^*_\textrm{vi}} = \frac{T+20}{T} $,
which matches the experimental observations. Consequently, to obtain an $ \epsilon $-suboptimal solution, $ 20 / \epsilon $ iterations are required.

When handling this relatively large network, the policy iteration approach \cite{wu2019optimal} runs out of memory during Jordan decomposition of huge matrices (see Eq. (18) in \cite{wu2019optimal}). This fact indicates the potentially high space complexity of the policy iteration approach. To compare the running time, we have to use a smaller network instead: the \textit{E. coil} Ara operon network considered in \cite{wu2019optimal}, which has 9 nodes and 4 inputs. It is reported in \cite{wu2019optimal} that the policy iteration approach takes 8.54~s with only three iterations to get the optimal solution. By contrast, our approach only needs about 0.14 s to get the same optimal value. Since our hardware capacity in this study  is similar to that in \cite{wu2019optimal}, the difference between running time demonstrates the superior time efficiency of our approach.

\begin{remark}  \label{rk: jordan}
	Though advanced numerical routines may be used to save memory for Jordan decomposition in the policy iteration algorithm \cite{wu2019optimal}, which is beyond the scope of this study, such matrix decomposition is much more complicated than the simple operations in our algorithms. \rv{Besides, we have tested these algorithms' performance with a variety of initial states, and the running time comparison remains unchanged: our algorithm always runs much faster.}
\end{remark}

\section{Conclusion} \label{sec: conclusion}
This paper dealt with the infinite-horizon optimal control (IHOC) problem of SBCNs with average cost from a graph-theoretical point of view. We built a graph structure, named the optimal state transition graph, to organize the reachable states and to determine the optimal switching-control pair for each one-step transition. The infinite-horizon problem was reduced to a minimum-mean cycle (MMC) problem in this graph, which was subsequently solved efficiently by adapting Karp's algorithm. Besides, we managed to design a static state-feedback control and switching law by picking the optimal trajectory wisely. Of course, we note that the solutions to Problem \ref{prob: 1} are generally not unique, and our approach yields a concise one that can be easily implemented by state feedback. Both time complexity analysis and a benchmarking test with the T-LGL network have confirmed the superior efficiency of our approach that makes it more scalable to relatively large networks, \rv{though it still runs in exponential time.}

\rv{Although we developed the optimal control method primarily for SBCNs in this paper, the proposed method can be potentially extended to more general logical networks. A common generalization of a BCN is a $ k $-valued logical network (KVLN), of which each variable has $ k > 2 $ possible values \cite{zhao2010optimal, cheng2015receding}. The extension of our approach to switched KVLNs is straightforward, since the ASSR of a KVLN is exactly the same as that of a BCN but with $ N \deq k^n $ and $ M \deq k^m $.  The constraints on states, control inputs, and switching actions can be handled similarly for KVLNs. By contrast, it seems unlikely that the proposed approach can be easily extended to PBNs and time-delayed BCNs \cite{zhu2019asymptotical}, because the state transition of the former is stochastic and, in the latter, an edge of the STG no longer represents a state transition attained in one
time step due to the delayed dynamics. Nonetheless, it is still meaningful to explore such possibilities following a graph-theoretical idea.
More generally, it deserves further investigations to examine other control-theoretical problems for BCNs and SBCNs by combining the ASSR and graph theory.}


%

\bibliographystyle{IEEEtran}
\bibliography{main_ac_cybernetics}

\begin{IEEEbiography}
	[{\includegraphics[width=1in,height=1.25in, clip,keepaspectratio]{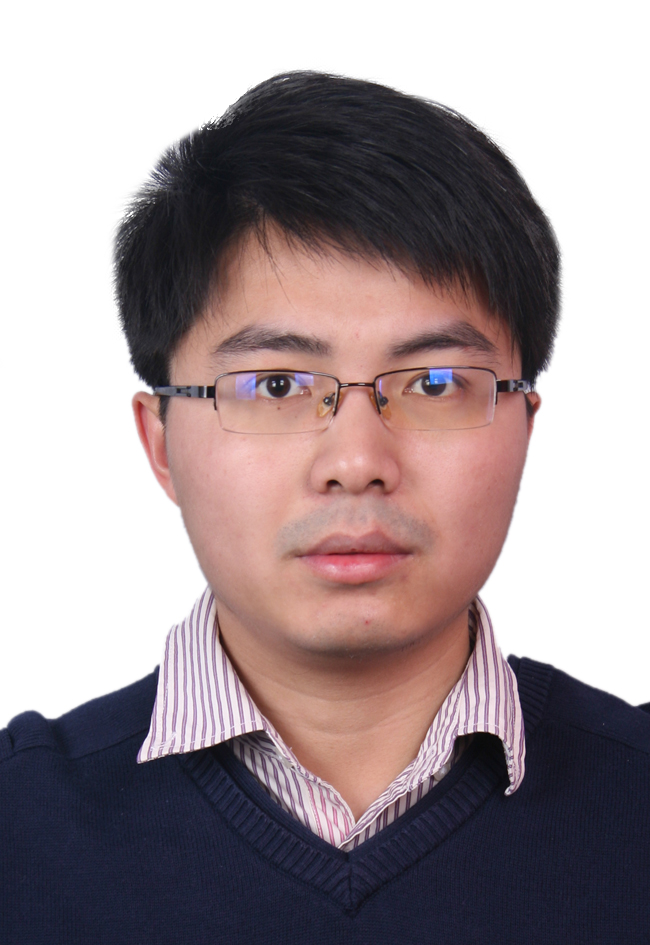}}]
	{Shuhua Gao} received the B.S. degree from Shanghai Jiao Tong University in 2012, the M.S. degree from Beihang University in 2015, both in mechanical engineering, and the Ph.D. degree in electrical and computer Engineering from National University of Singapore in 2020, where he is currently a research fellow.
	His research interests include Boolean networks in systems biology, optimal control, computational intelligence and its application in control engineering, and robotics. 
\end{IEEEbiography}

\begin{IEEEbiography}
	[{\includegraphics[width=1in,height=1.25in, clip,keepaspectratio]{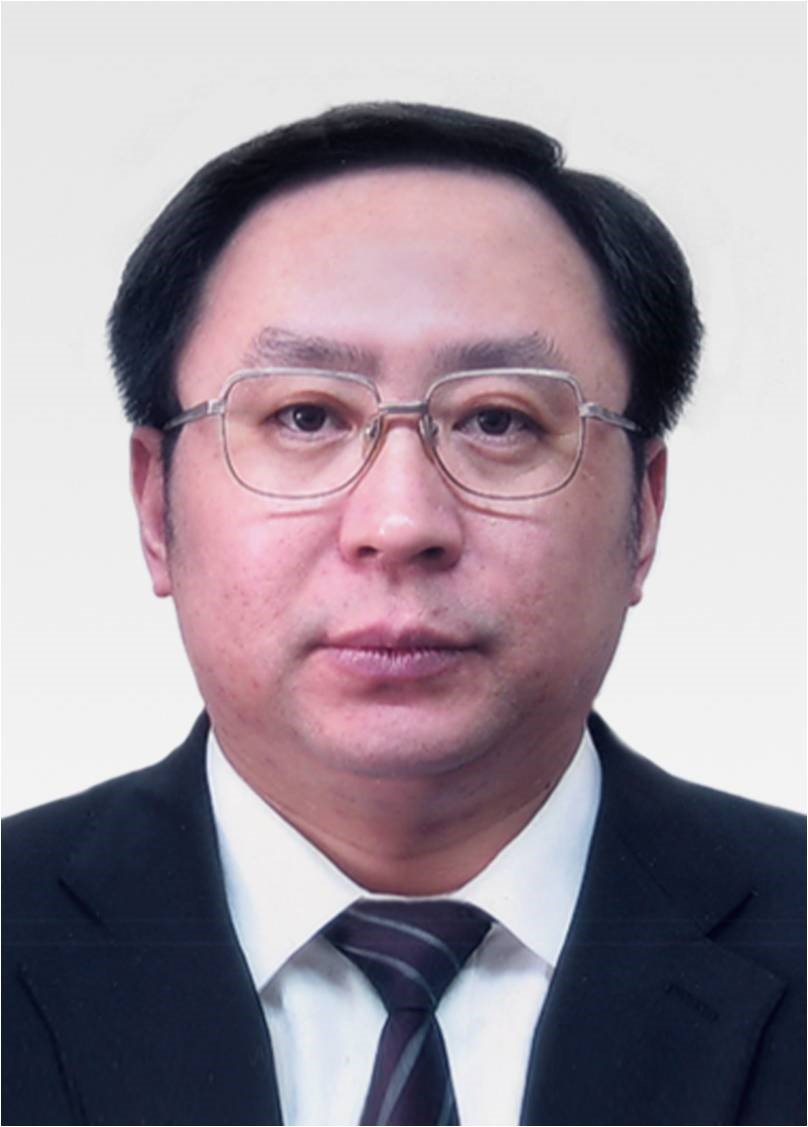}}]
	{Changkai Sun} received the B.S. degree in medicine from The Second Military Medical University, Shanghai, China, in 1986, and the Master degree in clinical neurology in 1995 and the M.D./Ph.D. degree in human anatomy in 1997 from The Fourth Military Medical University, Xi’an, China, respectively. He was a postdoc fellow at the Institute of Basic Medical Sciences, Academy of Military Medical Sciences, Beijing, China, from 2000 to 2003, and a visiting scholar with the Yue Lab of Neurophysiology, Institute of Biomedical Engineering, and Najm epilepsy center, Institute of Neurology, Cleveland Clinic, USA, from 2005 to 2007. He is currently the Chairman and a Professor with Research \& Educational Center for the Control Engineering of Translational Precision Medicine (RECCE-TPM), 
	School of Biomedical Engineering, Faculty of Electronic Information and Electrical Engineering, Dalian University of Technology, China. His current research interests include a multidisciplinary innovation of rBNN-rBNN$ ^+ $-iANN about real and artificial neural networks.
\end{IEEEbiography}

\begin{IEEEbiography}
[{\includegraphics[width=1in,height=1.25in, clip,keepaspectratio]{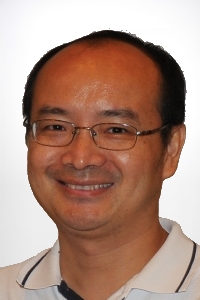}}]
{Cheng Xiang} received the B.S. degree from Fudan University in 1991; M.S. degree from the Institute of Mechanics, Chinese
Academy of Sciences in 1994; and M.S. and
Ph.D. degrees in electrical engineering from Yale
University in 1995 and 2000, respectively. He
is an Associate Professor and the area director of Control, Intelligent Systems \& Robotics in the Department of
Electrical and Computer Engineering at the National University of Singapore. His research interests include computational intelligence, adaptive systems, and pattern recognition.

\end{IEEEbiography}

\begin{IEEEbiography}
	[{\includegraphics[width=1in,height=1.25in, clip,keepaspectratio]{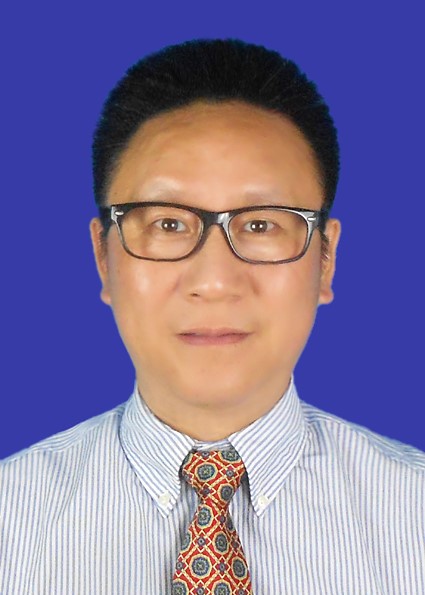}}]
	{Kairong Qin} received the B.S. degree in Mechanics from Fudan University, China, in 1991. He continued his study in Fluid Mechanics, Fudan University, China, and received the Ph.D. degree in 1996. He is currently a full professor in the School of Optoelectronic Engineering and Instrumentation Science, Dalian University of Technology, China. His research interests include biomechanics, mechanobiology, microfluidics and intelligent medical devices. 
\end{IEEEbiography}

\begin{IEEEbiography}
[{\includegraphics[width=1in,height=1.25in, clip,keepaspectratio]{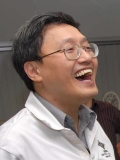}}]
{Tong Heng Lee} (M'90) received the B.A. degree
(First Class Hons.) in engineering tripos from
Cambridge University, Cambridge, U.K., in 1980,
and the Ph.D. degree in electrical engineering from
Yale University, New Haven, CT, USA, in 1987.
He is a Professor with the Department of
Electrical and Computer Engineering, National
University of Singapore. He is
the Deputy Editor-in-Chief of the IFAC Mechatronics journal and serves as the Associate
Editor of many other flagship journals. He has also coauthored three
research monographs and holds four patents. His research interests are in the areas of
adaptive control systems, knowledge-based control, mechatronics,
and computational intelligence. 
\end{IEEEbiography}
\end{document}